\documentclass[a4paper,twocolumn,11pt,accepted=2022-11-03]{quantumarticle}
\pdfoutput=1
\usepackage[utf8]{inputenc}
\usepackage[english]{babel}
\usepackage[T1]{fontenc}

\usepackage{lipsum}

\usepackage{amsmath, amsfonts, amssymb, amsthm, bbm}
\usepackage{dsfont}
\usepackage[T1]{fontenc}
\usepackage{enumitem}
\usepackage{bm}
\usepackage{mathtools}
\usepackage{bm}
\usepackage{hyperref}
\usepackage{physics}
\usepackage{listings}
\usepackage[table,xcdraw,rgb]{xcolor}

\usepackage{tikz}
\usepackage[numbers,sort&compress]{natbib}
\usetikzlibrary{shapes,backgrounds}
\definecolor{blueCB}{HTML}{68007D}
\definecolor{blueA}{HTML}{36CDFF}
\definecolor{blueB}{HTML}{959FC9}
\definecolor{blueC}{HTML}{D591E3}

\input{Qcircuit}

\begin{document}
\title{Efficient simulation of Gottesman-Kitaev-Preskill states with Gaussian circuits}

\author{Cameron Calcluth}
\affiliation{Department of Microtechnology and Nanoscience (MC2), Chalmers University of Technology, SE-412 96 G\"{o}teborg, Sweden}
\orcid{0000-0001-7654-9356}
\email{calcluth@gmail.com}
\author{Alessandro Ferraro}
\affiliation{Centre for Theoretical Atomic, Molecular and Optical Physics, Queen's University Belfast, Belfast BT7 1NN, United Kingdom}
\affiliation{Dipartimento di Fisica ``Aldo Pontremoli,''
Università degli Studi di Milano, I-20133 Milano, Italy}
\author{Giulia Ferrini}
\affiliation{Department of Microtechnology and Nanoscience (MC2), Chalmers University of Technology, SE-412 96 G\"{o}teborg, Sweden}

\maketitle

\begin{abstract}
We study the classical simulatability of Gottesman-Kitaev-Preskill (GKP) states in combination with arbitrary displacements, a large set of symplectic operations and homodyne measurements. For these types of circuits, neither continuous-variable theorems based on the non-negativity of quasi-probability distributions nor discrete-variable theorems such as the Gottesman-Knill theorem can be employed to assess the simulatability. We first develop a method to evaluate the probability density function corresponding to measuring a single GKP state in the position basis following arbitrary squeezing and a large set of rotations. This method involves evaluating a transformed Jacobi theta function using techniques from analytic number theory. We then use this result to identify two large classes of multimode circuits which are classically efficiently simulatable and are not contained by the GKP encoded Clifford group. Our results extend the set of circuits previously known to be classically efficiently simulatable.
\end{abstract}
\section{Introduction}
Identifying quantum computing architectures that are capable of yielding quantum advantage, in contrast to classically efficiently simulatable ones, is of paramount importance, both at the fundamental level, and to design useful quantum machines capable of surpassing classical capability for computing tasks~\cite{harrow2017}.

Quantum computing architectures based on bosonic fields, to which quadrature operators with a continuous-variable spectrum can be associated, are attracting growing interest as they allow for implementing resource-efficient quantum error correction with the use of bosonic codes~\cite{gottesman2001,grimsmo2020,Joshi_2021,Shruti_Arne-2021,hillmann2021performance, ofek2016, fluhmann2019, hu_quantum_2019, campagne-ibarcq2020, kudra2021robust}.  
It has recently been shown that concatenating bosonic codes with qubit-types of codes increases the threshold of tolerable error probability in quantum error-correction, with respect to the use of only qubit error correction~\cite{vuillot2019quantum, noh2020fault, noh2022}.

One of the most promising bosonic codes is the Gottesman-Kitaev-Preskill (GKP) code~\cite{gottesman2001, Shruti_Arne-2021, fluhmann2019, campagne-ibarcq2020, kudra2021robust}, which allows for correcting arbitrary errors. GKP codewords also have a highly negative Wigner function~\cite{gottesman2001,garcia-alvarez2019,yamasaki2020}, which is a known necessary and quantifiable
resource for universal quantum computation with continuous variables~\cite{mari2012,veitch2012}. Furthermore, GKP codewords corresponding to the 0-logical state have been shown to be universal in the resource-theoretic sense~\cite{albarelli2018, takagi2018} --- i.e., they promote a Gaussian set of operations, states and measurements, to fault-tolerant universal quantum computation~\cite{baragiola2019}. 

In contrast to these results, there are certain circuits involving GKP states which are efficiently simulatable. 
For instance, it is known that encoded stabilizer GKP states in combination with special Gaussian circuits, made of discrete displacements and encoded qubit and qudit Clifford operations, measured with homodyne detection, are classically efficiently simulatable~\cite{garcia-alvarez2020}. Hence, the separation between circuits involving GKP states which can be simulated on a classical computer and those which cannot remains unclear.

In this work, we tackle the unexplored question: are GKP states in combination with arbitrary displacements, Gaussian gates and homodyne measurement universal, or classically efficiently simulatable?
For these types of circuits, neither continuous-variable theorems based on the non-negativity of quasi-probability distributions nor discrete-variable theorems such as the Gottesman-Knill theorem can be employed to assess the simulatability.

Here, we prove that a large class of such Gaussian circuits is classically efficiently simulatable. The operations that we consider are a large subset of the Gaussian operations, which are not contained within the continuous-variable Clifford group and are therefore not simulatable by the previous methods used in Ref.~\cite{garcia-alvarez2020}. To highlight the distinction from these previous methods, we consider the case of single-mode and multimode measurements separately.

The circuits in Ref.~\cite{garcia-alvarez2020} are simulatable when the operations consist of multimode Gaussian operations whereby the parameters of the symplectic operations are selected from a zero-measure set and the parameters of the displacements are selected from the set of rational numbers. In this work, when restricting to the measurement of a single mode, we demonstrate simulatability for multimode operations whereby the parameters of symplectic operations are chosen from a set that is dense in the reals. Furthermore, for single-mode measurements, simulation can be achieved in time which increases linearly with respect to the number of modes. Meanwhile, we extend the class of simulatable displacements from rational to all possible continuous displacements.

Furthermore, for multimode measurement after multimode Gaussian operations, we demonstrate that another large set of symplectic operations and all real displacements are simulatable. None of these sets is contained within the set of operations given in Ref.~\cite{garcia-alvarez2020} and therefore this work further increases the known set of simulatable operations acting on $0$-logical GKP states.

The simulation methods given in this paper are distinct from both stabilizer methods~\cite{gottesman1999,bartlett2002,aaronson2004,van-den-nest2010,beaudrap2013,gheorghiu2014,veitch2014,Juani-thesis,bermejo-vega2014} and phase space methods~\cite{cahill1969,mari2012,veitch2012,rahimi-keshari2016}. Our method is based on deriving an expression for the probability density function (PDF) with respect to position for any Gaussian operation acting on any initial states.

When choosing $0$-logical GKP states as input and restricting the parameters of the symplectic operations to a set which is dense in the reals, this PDF can be evaluated efficiently. The evaluation of the total PDF relies on the ability to evaluate the PDF of individual rotated and squeezed $0$-logical GKP states, which we demonstrate is possible for certain rotation parameters. This calculation is performed by analyzing the rotated wave function and simplifying it using techniques from analytic number theory. In particular, we simplify a Jacobi theta function into an expression involving the quadratic Gauss sum which has a solution for specific parameters.

In Section \ref{sec:tracking-evolution} we present the general circuit class which we investigate throughout this work; we describe Gaussian operations and outline how to track the evolution of measurement operators in the Heisenberg picture. In Section \ref{sec:pdf-single} we present our first key result, i.e., an analytic expression for the PDF of a GKP state with respect to position, which has undergone an arbitrary Gaussian operation. Furthermore, we show that for angles of rotation selected from a set dense in the reals, the PDF can be reduced to a Dirac comb using theorems from analytic number theory. We use this result in Section \ref{sec:simulation-singlemode} to show our second major contribution, namely that a large set of multimode Gaussian operations followed by a single-mode measurement are efficiently classically simulatable. Our final main finding presented in Section \ref{sec:simulation-multimode} demonstrates that we can extend these results to multimode measurements on the condition that the multimode operations are chosen from a further restricted set of Gaussian operations. In Section \ref{sec:summarysimulatable} we summarize the sets of operations which are simulatable in both the single-mode measurement case and the multimode measurement case and compare them with discrete-variable simulation techniques. In Section \ref{sec:realistic-gkp} we provide an analytic expression for the rotated and squeezed realistic GKP state. We also provide a brief overview of the challenges of extending our results to the simulation of circuits involving realistic GKP states. Finally, our concluding remarks are presented in Section \ref{sec:conclusion}.

\section{Gaussian operations and basic formalism}
\label{sec:tracking-evolution}

In this Section we lay down the basic formalism required to tackle the circuits that we consider, namely we define the input GKP states, review the evolution under Gaussian operations, as well as recall the symplectic formalism. 

\subsection{The circuits considered}
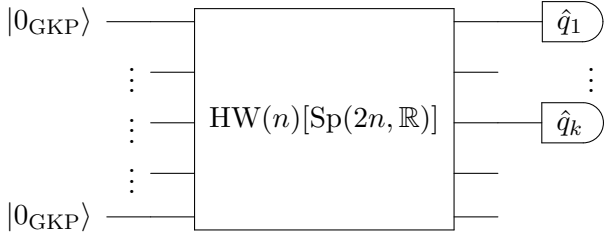
\begin{figure}[h!]
     \centering
     $$
     \Qcircuit @C=1.5em @R=0.6em {
     \push{\rule{1.2em}{0em}}&\lstick{\ket{0_{\text{GKP}}}} & \qw & \multigate{4}{ \text{HW}(n)[ \text{Sp}(2n,\mathbb R)]} & \qw & \measureD{\hat q_1}\\
     && \lstick{\vdots} & \ghost{ \text{HW}(n)[ \text{Sp}(2n,\mathbb R)]} & \qw & \rstick{\vdots} \\
     && \lstick{\vdots} & \ghost{ \text{HW}(n)[ \text{Sp}(2n,\mathbb R)]} & \qw & \measureD{\hat q_k}  \\
     && \lstick{\vdots} & \ghost{ \text{HW}(n)[ \text{Sp}(2n,\mathbb R)]} & \qw &  \\
     \push{\rule{1.2em}{0em}}&\lstick{\ket{0_{\text{GKP}}}}  & \qw & \ghost{ \text{HW}(n)[ \text{Sp}(2n,\mathbb R)]} & \qw &}
     $$
     \caption{Schematics of the general circuit class that is considered. The input states are $0$-logical GKP states. The states are acted on by generic Gaussian operations. Homodyne detection of $k$ modes follows, corresponding to the measurement of the quadratures $\hat q_1,\dots \hat q_k$.}
     \label{fig:allgaussian-circuits}
    \end{figure}

We consider bosonic systems consisting of an arbitrary number $n$ of bosonic modes, each characterized by canonical position $\hat q_k$ and momentum $\hat p_k$ operators, with $[\hat q_k, \hat p_l] = i \delta_{k,l}$.
In particular, we focus on  the class of circuits schematized in Fig.~\ref{fig:allgaussian-circuits}.

The input states are $0$-logical square GKP states\footnote{The GKP kets, as defined in Eq.~(\ref{eq:gkp-kets}), do not formally represent proper quantum states as they are not normalizable, and as such do not belong to the Hilbert space associated with a bosonic mode. However, in the remainder of this paper, we will nonetheless adopt the colloquial expression GKP states, for the sake of brevity.} which have a wave function in position representation given by~\cite{gottesman2001}
    \begin{align}
        \label{eq:gkp-kets}
        \psi_{0,L}(x)=\bra{q=x}\ket{0_{\text{GKP}}}=\sum_m \delta(2m\sqrt\pi - x);
    \end{align}
    all in all, the input state can hence be compactly indicated by the tensor product over all the $n$ input modes initialized in the $0$-logical GKP state, $ \ket{\mathbf 0_{\text{GKP}}}=\otimes_{j=1}^n \ket{0_{\text{GKP}}}_j$.
    The density matrix of the initial state is given by
    \begin{align}
    \label{eq:input-state-density}
        {\hat \rho_0=\ket{\mathbf 0_{\text{GKP}}}\bra{\mathbf 0_{\text{GKP}}}}.
    \end{align}

The evolutions that we consider are a subset of all Gaussian operations on $n$ modes. We will define the class considered in Sections \ref{sec:simulation-singlemode} and \ref{sec:simulation-multimode}. Measurement occurs through homodyne detection, say corresponding to the measurement of the position quadratures $\hat q$ of a single mode (Section \ref{sec:simulation-singlemode}) or all modes (Section \ref{sec:simulation-multimode}). In the following, we review the evolution of the quadrature operators under Gaussian evolutions which we will later use to determine the PDF of Gaussian circuits with input GKP states.

\subsection{Tracking evolutions under Gaussian operations}
Analogously to the Clifford group in discrete-variable quantum computation, Gaussian operations preserve the continuous-variable Pauli (or Heisenberg-Weyl) group $\text{HW}(n)$, which is the group generated by the set $\{e^{ic_j\hat p_j},e^{-id_j \hat q_j}~:~j~\in~\mathbb{Z}_n\quad c_j,d_j\in\mathbb{R}\}$~\cite{bartlett2002}. They can be constructed by selecting and applying any number of the following generators in any order~\cite{bartlett2002,braunstein:05, gu2009}:
\begin{equation}
\label{eq:generator-Gaussian}
    \{e^{ic_j\hat q_j},e^{i \theta_j(\hat q_j^2 + \hat p_j^2)/2},e^{-i\ln{s_j}(\hat q_j\hat p_j+\hat p_j\hat q_j)/2},e^{-i\hat q_j\hat p_k}\}
\end{equation}
for ${c_j,s_j\in \mathbb R}$, $\theta_j\in [0,2\pi) \in\mathbb R$ and ${j,k\in \{1,\dots, n\}}$.

Our approach to tackle the computation of the PDF of GKP states after Gaussian operations starts with  computing the Heisenberg evolution of the position operators $\hat q_j$ and their conjugate momentum operators $\hat p_j$  under the action of a unitary operator $\hat U$. The generator gates in Eq.~(\ref{eq:generator-Gaussian}) have the effect of transforming the quadrature operators $\hat q_j$ and $\hat p_j$ linearly.
Displacements $e^{i c_j \hat q_j}$ have a non-trivial effect only on the momentum of the $j$-th mode, given by
\begin{align}
    \hat p_j \rightarrow \hat p_j +c_j.
\end{align}
Similarly rotations $\hat R(\theta_j)=e^{i\theta_j(\hat q_j^2+\hat p_j^2)/2}$ have the effect 
\begin{align}
    \hat q_j &\rightarrow \hat q_j \cos\theta_j-\hat p_j\sin\theta_j,\nonumber\\
    \hat p_j &\rightarrow \hat q_j\sin\theta_j+\hat p_j\cos\theta_j.
\end{align}
The Fourier transform can be defined as the rotation with angle $\pi/2$, i.e. ${\hat F=\hat R(\pi/2)=e^{i\pi(\hat q_j^2+\hat p_j^2)/4}}$, which has the effect
\begin{align}
    \label{eq:fourier-effects}
    \hat q_j\to -\hat p_j, \quad  \hat p_j \to \hat q_j.
\end{align}
The Fourier transform acting on encoded GKP qubits corresponds to the Hadamard gate. Squeezing operations ${\hat S(s_j)=e^{-i\ln{s_j}(\hat q_j\hat p_j+\hat p_j\hat q_j)/2}}$ have the effect
\begin{align}
    \hat q_j \rightarrow s_j \hat q_j, \quad \hat p_j \rightarrow \hat p_j/s_j.
\end{align}
Finally, the SUM gate $\hat C_X=e^{-i\hat q_j\hat p_k}$, where $j$ is the control mode and $k$ is the target mode, has the effect
\begin{alignat}{2}
    \label{eq:sumgate-effects}
    \hat q_j&\rightarrow \hat q_j,  && \hat p_j\rightarrow \hat p_j-\hat p_k\nonumber\\
    \hat q_k&\rightarrow \hat q_j+\hat q_k,\quad     &&\hat p_k\rightarrow \hat p_k.
\end{alignat}

As can be seen by these Heisenberg evolutions, we can track the evolution of the measurement operator associated with the modes of interest using a linear equation of the form
    \begin{align}
    \label{eq:evolved-quadrature}
        \hat Q_j = \hat U^\dagger \hat q_j\hat U=\sum_ia^{(j)}_{i} \hat q_i + b^{(j)}_{i} \hat p_i+ c_j,
    \end{align}
{\it i.e.}, we need to track $2n+1$ real variables for each mode we would like to measure. To measure $n$ modes we need to track $(2n+1)n$ real variables. Similarly to Ref.~\cite{bartlett2002}, for any finite precision, this can be done efficiently as long as the number of gates is polynomial in the number of modes.

\subsection{Decomposition of Gaussian operations}
We will now analyze the effect of any quadratic Hamiltonian on the measurement mode $\hat q_j$. We first introduce the vector of operators~\cite{ferraro2005, serafini2017}
\begin{align}
    \hat{\mathbf r}=\mqty(\hat q_1,\dots,\hat q_n,\hat p_1,\dots,\hat p_n)^T
\end{align}
and the group of symplectic matrices
\begin{align}
    \label{eq:symplecticset-main}
    \text{Sp}(2n,\mathbb R)=\{M\in M_{2n\times 2n}(\mathbb R):M^{\mathrm {T} }\Omega M=\Omega \},
\end{align}
where
\begin{align}
    \Omega=\mqty(0&-\mathbbm{1}_n\\\mathbbm{1}_n&0).
\end{align}
Any unitary operator generated by a quadratic Hamiltonian $\hat{H}= \hat{\mathbf r}^TH\hat{\mathbf r}/2$ can be associated to a symplectic matrix $M$ as 
\begin{align}
    \hat U_M=e^{\frac i 2 \hat{\mathbf r}^TH\hat{\mathbf r}} , \quad  \text{ where } \quad M=e^{-\Omega H}.
\end{align}
This operation acting on the operators ${\hat{r}_j\to\hat U_M^\dagger \hat r_j \hat U_M}$ can be expressed in terms of a mapping of the vector of operators 
\begin{align}
    \hat{\mathbf r}\to M\hat{\mathbf r}.
\end{align}
Any linear displacement can be achieved using an operator
\begin{align}
    \hat D_{\bar{\mathbf r}}=e^{-i\hat{\mathbf r}^T\Omega \bar{\mathbf r}}
\end{align}
parameterized by a vector of $2n$ real numbers $\bar{\mathbf r}$ corresponding to the magnitude of displacement in position and momentum. Its effect $\hat D_{\bar{\mathbf r}}^\dagger \hat{\mathbf r}\hat D_{\bar{\mathbf r}}$ on the vector of operators can be expressed as
\begin{align}
    \hat{\mathbf r}\to \hat{\mathbf r}+\bar{\mathbf r}.
\end{align}

Utilizing this formalism, unitary operations corresponding to any quadratic Hamiltonian can always be decomposed as a Gaussian operation of the form
\begin{align}
    \label{eq:defineU}
    \hat U=\hat U(\bar{\mathbf r},M)=\hat D_{\bar{\mathbf r}}\hat U_M
\end{align}
which acts on the quadratures $\hat{\mathbf r}$ as
\begin{align}
    \hat U^\dagger \hat{\mathbf r} \hat U=M\hat{\mathbf r}+\bar{\mathbf r}.
\end{align}
We will denote $\hat U(\bar{\mathbf r},M)=\hat U$ for simplicity, and will use $\hat U$ throughout the rest of this work.
The matrix $M$ can be written in block form as
\begin{align}
    \label{eq:block-form}
    M=\mqty(A&B\\C&D).
\end{align}
The effect of $\hat U$ on the Heisenberg measurement modes can be evaluated as linear equations as in Eq.~(\ref{eq:evolved-quadrature}).

Therefore, we see that any quantum computation composed of Gaussian operations in any order followed by measurements in $\hat q_j$ can be decomposed in terms of a symplectic operation, followed by linear displacements in position with parameters
\begin{align}
    r_i = c_i \quad \forall \quad  i\in\{n+1,\dots,2n\}
\end{align}
and the block matrices
\begin{align}
    A_{i,j}=a^{(j)}_i \quad B_{i,j}=b^{(j)}_i.
\end{align}
Note that the values $r_i$ for $i\le n$ can be any real value since they correspond to displacements in momentum, which commutes with the position measurement operator.

In the following sections, we will show how to simulate a restricted set of Gaussian operations acting on $0$-logical GKP states, followed by measurements in $\hat q_j$.
\section{PDF of a Gaussian-transformed GKP state}
\label{sec:pdf-single}
The PDF with respect to position of a $0$-logical GKP state following a general single-mode Gaussian operation consists of a Gaussian operator applied to an infinite sum of Dirac delta peaks. The Gaussian operator will introduce non-trivial coefficients at each of these peaks. It is known that for a restricted set of Gaussian operations --- e.g., those which can be described in terms of GKP-encoded single-qubit Clifford operations --- it is possible to analytically evaluate the wave function, and therefore also the PDF of a transformed GKP state~\cite{gottesman2001}. However, to the best of our knowledge, no general result is known for arbitrary Gaussian operations.

In this section, we provide the evaluation of the PDF of a $0$-logical GKP state that has undergone a generic Gaussian transformation --- which generally cannot be described in terms of GKP-encoded qubit Clifford operations --- in the position basis. We note that we can decompose any single-mode symplectic operation using the Iwasawa decomposition~\cite{arvind1995} in terms of a shear operation followed by a squeezing operation and then a rotation. Furthermore, when measuring in the position basis, we can discard the initial shear operation, as it has no effect on the position quadrature. Therefore, the effect of all single-mode Gaussian operations acting on the position quadrature can be expressed in terms of a displacement, squeezing and a rotation. Since the displacement has the effect of a displacement of the position variable in the PDF, we can, without loss of generality, ignore the effect of displacement in the following calculation.

The wave function of the rotated and squeezed $0$-logical GKP state can be expressed in terms of these single-mode operations as
\begin{align}
    \psi_{\theta,s}(x)&=\bra{q=x}\hat S(s)
    \hat R(\theta)\ket{0_{\text{GKP}}}.
\end{align}
Furthermore, we can isolate the squeezing parameter $s$, which has the effect of rescaling the wave function of a rotated 0-logical GKP state,
\begin{align}
    \psi_{\theta,s}(x)=\psi_\theta(x/s).
    \label{eq:justtheta}
\end{align}
The full calculations detailed in Appendix \ref{sec:appendix-singlemodeops-transforming} show that the PDF of a rotated 0-logical GKP state measured in the position basis can be written compactly as being proportional to
\begin{align}
    \abs{\psi_{\theta}(x)}^2\propto&\frac{1}{ \sin\theta} \abs{\vartheta(\zeta=-x \csc\theta /\sqrt\pi;\tau=2\cot\theta)}^2,
\end{align}
where the Jacobi theta function $\vartheta(\zeta;\tau)$ is defined as
\begin{align}
    \label{eq:jacobi}
    \vartheta(\zeta;\tau)\equiv\sum_{m\in \mathbb Z}e^{\pi i m^2 \tau}e^{2\pi i m \zeta }.
\end{align}
We consider two cases for the rotation angle $\theta$, 
for which the PDF can be evaluated analytically.
For case 1 we consider rotation angles $\theta$ for which ${\cot\theta=u/v}$ where ${u\in\mathbb Z,v\in\mathbb Z_{\text{odd}}}$. For case 2 we consider rotations angles where $\theta\mod \pi=0$.
These cases correspond to selecting $\theta\in\Theta$ where the set $\Theta$ is defined as
\begin{align}
    \label{eq:anglesset}
    \Theta=\{\theta \in \mathbb R: \cot\theta =u/v \in \mathbb Q_{(2)} \} \cup \{0,\pi\}
\end{align}
and $\mathbb Q_{(2)}$ is formally defined as the localization of the integers $\mathbb Z$ at the prime ideal $2\mathbb Z$~\cite{atiyah2019}.

First, we summarize the calculations of case 1, where the rotation angles can be expressed as ${\cot\theta=u/v}$. By expanding the theta function in terms of $u,v$, it is possible to express it in terms of a Dirac comb and a quadratic Gauss sum. The Gauss sum is given by
\begin{align}
    G(\zeta, \tau=2u/v)=\sum_{m\in \mathbb Z_{v}} e^{2\pi i m^2 u/v}e^{2\pi i m \zeta},
\end{align}
and the theta function can be written in terms of the Gauss sum as
\begin{align}
\label{eq:thetaandgauss-main-original}
    \vartheta(\zeta;\tau=2u/v)=&G(\zeta,\tau)\sum_{n'} \delta(\zeta-n'/v).
\end{align}
The delta function evaluates to $0$ except at values of $\zeta$ equal to $n'/v$ so the theta function can be simplified to
\begin{align}
\label{eq:thetaandgauss-main}
    \vartheta(\zeta;\tau=2u/v)=&G(\zeta=n'/v,\tau)\sum_{n'} \delta(\zeta-n'/v).
\end{align}
Using results from analytic number theory we identify that for the case of $\theta$ we consider, the Gauss sum evaluates to a constant. This implies that the norm of the theta function is proportional to a Dirac comb. Hence, the PDF for this case can be evaluated to
\begin{align}
    |\psi_{\theta}(x)|^2
    =&\sum_{m} \delta( x -m\sqrt\pi\sin\theta/v),
\end{align}
where we have ignored the normalization constant as in Ref.~\cite{gottesman2001}. Formally, this PDF is not normalizable, however, we can still interpret it as a probability density function in that it informs us that the measured position value $x$ will be given by any integer $m$ multiple of $\sqrt\pi\sin\theta/v$ with equal probability.

In the second case, we identify that a rotation of ${\theta=0}$ gives the identity, and a rotation by ${\theta=\pi}$ is equivalent to applying the Fourier transform twice, which is also the identity. Therefore, we can immediately write the PDF as
\begin{align}
    \abs{\psi_{\theta}(x)}^2
    = &\sum_m\delta(x-2m\sqrt\pi).
\end{align}

We can unify both of these cases in a single PDF,
\begin{align}
\label{eq:rotated-wfs-main}
\abs{\psi_{\theta}(x)}^2=\sum_{m}\delta(x-m \sqrt\pi  \Delta),
\end{align}
with separation between peaks $\Delta$ as calculated in Appendix \ref{sec:appendix-singlemodeops-separation}, yielding
\begin{align}
\label{eq:deltacases-main}
    \Delta=\begin{cases}
    \sin\theta/v\quad & \text{ if } \cot\theta=u/v: u\in\mathbb Z,v\in\mathbb Z_{\text{odd}},\\
    2\quad &\text{ if } \theta=k\pi \text{ for } k \in \mathbb Z
    \end{cases}
\end{align}
where $\gcd(u,v)=1$. The PDF of the rotated and squeezed mode can therefore be expressed as
\begin{align}
\label{eq:singlemodepdf-main}
\abs{\psi_{\theta,s}(x)}^2=\sum_{m}\delta(x-m \sqrt\pi  s \Delta ),
\end{align}
where we again ignore the normalization constant.
    
\section{Simulatability of Gaussian circuits with single-mode measurement}
\label{sec:simulation-singlemode}

\begin{figure}[!h]
     \centering
     $$
     \Qcircuit @C=1em @R=0.6em {
     \push{\rule{1.8em}{0em}}&\lstick{\ket{0_{\text{GKP}}}} & \qw & \multigate{2}{\text{HW}(n)\times \text{RSp}(2n,\mathbb R)} & \qw & \measureD{\hat q_1}\\
     && \lstick{\vdots} & \ghost{\text{HW}(n)\times \text{RSp}(2n,\mathbb R)} & \qw &  \\
     \push{\rule{1.8em}{0em}}&\lstick{\ket{0_{\text{GKP}}}}  & \qw & \ghost{\text{HW}(n)\times \text{RSp}(2n,\mathbb R)} & \qw &}
     $$
     \caption{Schematics of the circuit considered in Section \ref{sec:simulation-singlemode}. The circuit is initialized in $0$-logical GKP states. The operations considered are in ${\text{HW}(n)\times \text{RSp}(2n,\mathbb R)}$, which is a restricted set of Gaussian operations. Homodyne detection follows, corresponding e.g. to the measurement of the quadrature $\hat q_1$.}
     \label{fig:singlemodecircuitclass-maintext}
    \end{figure}
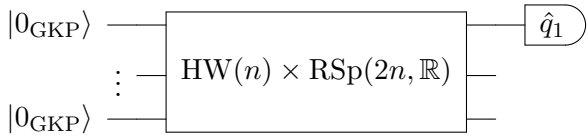
    For a restricted class of the circuits shown in Fig.~\ref{fig:allgaussian-circuits}, where the restriction arises from the two cases of allowed angles in Eq.~(\ref{eq:anglesset}), it is possible to simulate the outcome of the measurement. Specifically, we can consider a restricted class of multimode Gaussian operations $\mathcal A$ acting on $0$-logical GKP states followed by measurement of a single mode in the position quadrature, see Fig.~\ref{fig:singlemodecircuitclass-maintext}.

We wish to establish whether the PDF of the corresponding measurement outcomes
    \begin{align}
    \label{eq:probability-density}
        \text{PDF}(x_1)=&
        \Tr(\hat \rho \ket{x_1}\bra{x_1})\nonumber\\
        =& \Tr(\hat \rho_0\hat U^\dagger  \ket{x_1}\bra{x_1}\hat U ),
    \end{align}
    where $x_1$ are the eigenvalues of the position operator on the first mode $\hat q_1 \ket{x_1} =x_1 \ket{x_1}$ and $\hat \rho=\hat U\hat \rho_0\hat U^\dagger$,  can be calculated efficiently classically. 
This refers to the notion of strong simulatability, whereby the PDF of the outcomes is calculated in a time which scales at most polynomially both with the number of modes and the number of digits of precision~\cite{jozsa2014,terhal2004,bremner2010}.

We begin by detailing the set of Gaussian operations $\mathcal A$ that are simulatable by our technique. We will then show that the PDF of the multimode operations in $\mathcal A$ can be decomposed into the evaluation of an integral involving functions which correspond to the single-mode PDF of a rotated and squeezed $0$-logical GKP state, derived in Section \ref{sec:pdf-single}. The restricted set of multimode Gaussian operations we consider ensures that each single-mode PDF satisfies the constraints defined in the previous section.

The restricted class of operations that we consider, defined as
\begin{align}
    \label{eq:classA}
    \mathcal{A}=\text{HW}(n)\times\text{RSp}(2n,\mathbb R),
\end{align}
is the direct product\footnote{Note that in the definition of the Gaussian operations, the semi-direct product of the Heisenberg-Weyl group and the symplectic operations provides a complete description of all possible Gaussian operations since both the set of Heisenberg-Weyl operations and the set of symplectic operations are groups~\cite{bartlett2002}. However, in the restricted class of circuits we must use the direct product $\times$ since the restricted set of symplectic operations is not a group. This is equivalent to saying that the operations are selected from any combination of the two sets.} of the Heisenberg-Weyl group of all real displacements in phase space and the set of symplectic matrices ${M\in\text{RSp}(2n,\mathbb R)}$. The set $\text{RSp}(2n,\mathbb R)$ consists of all elements ${M\in\text{Sp}(2n,\mathbb R)}$ such that one of the following cases is satisfied
\begin{align}
    \label{eq:rspconstraints-maintext}
    A_{1,i} = 0 \text{ or }
    B_{1,i} = 0 \text { or }
    A_{1,i}/B_{1,i} =u_i/v_i
\end{align}
with $u_i\in\mathbb Z$ and $v_i\in\mathbb Z_{\text{odd}}$ for each $i\in\{1,\dots,n\}$, where $A,B$ are the block matrices of the symplectic matrix $M$ as defined in Eq.~(\ref{eq:block-form}).

 In other words, the class $\mathcal A$ contains operations of the form of $\hat U$ in Eq.~(\ref{eq:defineU}) where $(\bar{\mathbf r},M)$ is selected from the set of all possible phase space displacements $\bar{\mathbf r}\in\text{HW}(n)$ and symplectic operations parameterized by the symplectic matrix $M$ are such that the matrix satisfies the constraints in Eq.~(\ref{eq:rspconstraints-maintext}). We would like to demonstrate that when the operations $\hat U$ are selected from the class $\mathcal A$, acting on $0$-logical GKP states, followed by single-mode homodyne measurement of $\hat q$, the circuit is simulatable.

To calculate the PDF of the single-mode measurement circuit which is shown in Fig.~\ref{fig:singlemodecircuitclass-maintext}, we first track the evolution of the position quadrature in the Heisenberg picture, as in Eq.~(\ref{eq:evolved-quadrature}), which involves a summation over the $n$ modes of terms of the form
\begin{equation}
\label{eq:blocks-quadrature}
    a_{i} \hat q_i +b_{i} \hat p_i=s_{i}\left(\hat q_i \cos\theta_{i} - \hat p_i\sin\theta_{i}\right)=s_i\hat z_i^{\theta_i}
\end{equation}
where we have introduced the re-scaled parameters $s_i$ and $\theta_i$ such that ${a_{i} =s_{i}\cos\theta_{i}}$ and ${b_{i} =- s_{i}\sin\theta_{i}}$, and where we have defined the rotated quadrature $\hat z_i$. This is equivalent to performing a squeezing $\hat S(s_i)$ operation, followed by a rotation $\hat R(\theta_i)$, where $s_i\in\mathbb R$ and $\theta_i\in[0,2\pi)$. The coefficients $a_i,b_i,s_i,\theta_i$ depend on the mode tracked but we will omit the mode index in this section because we are only tracking a single mode, which we chose to be the first mode without loss of generality.

By evolving the measurement operator, rather than the states, we can express the PDF in Eq.~(\ref{eq:probability-density}) as
\begin{align}
    \label{eq:probability-density-heisenberg}
    \text{PDF}(x_1)= \Tr(\hat \rho_0  \ket{\hat Q_1=x_1}\bra{\hat Q_1=x_1}),
\end{align}
where the Heisenberg measurement operator can be calculated according to Eq.~(\ref{eq:evolved-quadrature}) and $\hat \rho_0$ is given in terms of Eq.~(\ref{eq:input-state-density}).
Using Eq.~(\ref{eq:blocks-quadrature}) we can then write each term of the sum as the Heisenberg operator for a transformed mode,
\begin{align}
    \label{eq:heisenberg-operator-single}
    \hat Q_1 = \sum_i s_i\hat z_i^{\theta_i} + c,
\end{align}
which allows us to calculate the PDF of the measurement of $\hat Q_1$. We can express the projection operator as a Dirac delta function, which allows us to write the PDF as
\begin{align}
    \text{PDF}(x_1)=&\bra{\mathbf 0_{\text{GKP}}}\delta\left(\hat Q_1-x_1\right)\ket{\mathbf 0_{\text{GKP}}}\nonumber\\
    =&\bra{\mathbf 0_{\text{GKP}}}\delta\left(\sum_i s_i\hat z_i^{\theta_i} + c-x_1\right)\ket{\mathbf 0_{\text{GKP}}}.
\end{align}
This can be evaluated by inserting the identity over all the transformed modes,
\begin{align}
    \label{eq:identity-rotated-modes}
    \mathbbm 1 =  \int \dd \mathbf z \prod_j^n\ket{\hat z_j^{\theta_j}=z_j}\bra{\hat z_j^{\theta_j}=z_j},
\end{align}
where $\mathbf z$ is the $n$-vector of the integration variables $z_j$.
This provides a PDF in terms of the wave functions of the individual transformed modes,
\begin{align}
        &\text{PDF}(\hat Q_1=x_1)\nonumber\\
        =& \int \dd\mathbf z \delta\left(\sum_{i=1}^{n} s_iz_i+c-x_1\right)\prod_j^n
        \abs{\psi_{\theta_j}(z_j)}^2,
\end{align}
where we have identified the PDF of individual transformed modes as
\begin{align}
    \abs{\psi_{\theta_j}(z_j)}^2=\abs{\bra{\hat z_j^{\theta_j}=z_j}\ket{0_{\text{GKP}}}}^2.
\end{align}

Inserting the PDF of the individual modes given by Eq.~(\ref{eq:singlemodepdf-main}) we find that the PDF of the measurement can be written compactly in terms of the parameters $s_i$ and $\Delta_i$ derived from the original symplectic matrix,
\begin{align}
    \label{eq:multimode-singlemeasure-pdf-main}
    &\text{PDF}(x_1) \nonumber\\
    =&
    \sum_{m_1,\dots,m_n\in\mathbb Z}\delta\left(\sum_{i=1}^{n}s_i m_i\sqrt\pi \Delta_i+c-x_1\right).
\end{align}
Note that this is possible when the symplectic matrix satisfies Eq.~(\ref{eq:rspconstraints-maintext}), in virtue of Section \ref{sec:pdf-single}. Further calculation details are provided in Appendix \ref{sec:appendix-singlemode-pdf}. The simulation of sampling from this circuit can be performed by generating $n$ random integers $m_i$ and returning
\begin{align}
    x_1=\sum_{i=1}^{n}s_i m_i\sqrt\pi \Delta_i+c.
\end{align}

The Gaussian operations selected from the class $\mathcal A$ can be parameterized in terms of the symplectic matrix elements $A_{1,i}$ and $B_{1,i}$. From these variables, we can calculate each $\Delta_i$ using arithmetic operations and the greatest common divisor following Eq.~(\ref{eq:deltacases-main}) and Eq.~(\ref{eq:rspconstraints-maintext}). The squeezing parameters $s_i$ can be evaluated with trigonometry from Eq.~(\ref{eq:blocks-quadrature}).
Furthermore, each of the variables $\Delta_i,s_i$ can be calculated independently. Therefore, the total number of these independent calculations scales linearly with the number of modes $n$.
Hence, we conclude that the PDF of the circuit can be calculated in linear time with respect to the number of modes. Therefore, the strong simulation of the class of circuits shown in Fig.~(\ref{fig:multimodecircuitclass-maintext}) can be performed in linear time and hence exists within complexity class $\text{P}$. 

For an example of a multimode circuit, involving the SUM and Fourier transform gates, which is simulatable by our technique, refer to Appendix \ref{sec:appendix-singlemode-example}. A circuit diagram with an explicit construction of a general circuit is given in Appendix \ref{sec:appendix-circuit-diagrams-A}.

These results are analogous to an input-GKP version of the results of Ref.~\cite{bartlett2002}, which show simulatability for Gaussian circuits and Gaussian measurements for position eigenstates. However, we stress that the proof techniques used for the derivation of our results are different and that the input GKP states are non-Gaussian and highly Wigner negative~\cite{gottesman2001,garcia-alvarez2019,yamasaki2020}, in contrast to the Gaussian input states of Ref.~\cite{bartlett2002}.

Practical tools for simulating Gaussian circuits with input GKP states were developed in Ref.~\cite{bourassa2021fast}, based on representing states as linear combinations of Gaussian states. Note however that their method is not computationally efficient for the circuits considered here, unlike ours. Tracking the evolution of the field quadratures is also used in the context of noise propagation under the twirling approximation to determine the fault-tolerance threshold for surface-GKP codes~\cite{noh2020fault,larsen2021faulttolerant, noh2022}.

Finally, note that the $0$-logical GKP state is possible to simulate using our technique due to the fact that its wave function can be written as a Dirac comb which does not have any position-dependent coefficients. We could also use other non-magic GKP states as input. For example, the state $\ket{1_{\text{GKP}}}$ and the encoded $X_L$ basis state $\ket{+_{\text{GKP}}}$ both satisfy the property that the peaks do not have position-dependent coefficients. Note that the encoded $X_L$ basis state $\ket{-_{\text{GKP}}}$ has an alternating sign for each peak, however, it can be generated using a momentum displacement applied to $\ket{+_{\text{GKP}}}$ and therefore is simulatable. If we instead consider the magic T-logical GKP state,
\begin{align}
    \ket{\text{T}_{\text{GKP}}}=\ket{0_{\text{GKP}}}+e^{i\pi/4}\ket{1_{\text{GKP}}},
\end{align}
simulation by our technique fails due to the complex term in the infinite summation over Dirac delta peaks. Simulation of this state would be surprising because it is considered resourceful in both continuous-variable and discrete-variable quantum computing. T-logical GKP states in combination with encoded Clifford operations are indeed hard to sample~\cite{yoganathan2019,garcia-alvarez2020}.

\section{Simulatability of Gaussian circuits with multimode measurement}
\label{sec:simulation-multimode}

\begin{figure}[h!]
     \centering
     $$
     \Qcircuit @C=1em @R=0.6em {
     \push{\rule{1.8em}{0em}}&\lstick{\ket{0_{\text{GKP}}}} & \qw & \multigate{2}{\text{HW}(n)\times \text{DSp}(2n,\mathbb R)} & \qw & \measureD{\hat q_1}\\
     && \lstick{\vdots} & \ghost{\text{HW}(n)\times \text{DSp}(2n,\mathbb R)} &  \rstick{\vdots} \qw\\
     \push{\rule{1.8em}{0em}}&\lstick{\ket{0_{\text{GKP}}}}  & \qw & \ghost{\text{HW}(n)\times \text{DSp}(2n,\mathbb R)} & \qw & \measureD{\hat q_n}}
     $$
     \caption{Schematics of the multimode measurement circuits considered. As in Fig.~\ref{fig:singlemodecircuitclass-maintext}, the input states are $0$-logical GKP stabilizer states. The operations considered belong to a further reduced set of Gaussian operations. Homodyne detection of multiple modes follows, corresponding e.g. to the measurement of the quadratures $\hat q_1,\dots,\hat q_n$.}
     \label{fig:multimodecircuitclass-maintext}
    \end{figure}
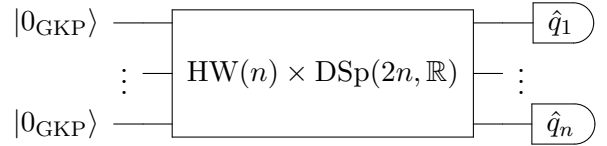
If we make an additional restriction to the set of symplectic matrices that we wish to simulate, we can generalize the results of the previous section to multimode measurement. We will denote this set of matrices $\text{DSp}(2n,\mathbb R)$. The set includes matrices of the form Eq.~(\ref{eq:block-form}) which can be decomposed to
\begin{align}
    \label{eq:operations-dsp}
    \mqty({\tilde A}&0\\{\tilde C}&({\tilde A}^{T})^{-1})\mqty(\text{diag}(\cos\vec \theta)&\text{diag}(\sin\vec \theta)\\
    -\text{diag}(\sin\vec \theta)&\text{diag}(\cos\vec \theta))
\end{align}
where ${\tilde A},{\tilde C}$ are $n\times n$ matrices such that ${\tilde A}$ is non-singular and both ${\tilde A}$ and ${\tilde C}^T{\tilde A}$ are symmetric\footnote{A $2n$-dimensional symplectic matrix can always be decomposed as~\cite{arvind1995}
\begin{align*}
S=\mqty(A&B\\C&D)=&\mqty(1&0\\{\tilde C}{\tilde A}^{-1}&1)\mqty({\tilde A}&0\\0&{\tilde A}^{-1})\mqty(X&Y\\-Y&X)\\
=&\mqty({\tilde A}&0\\{\tilde C}&{\tilde A}^{-1})\mqty(X&Y\\-Y&X)\\
=&\mqty({\tilde A}X&{\tilde A}Y\\{\tilde C}X-{\tilde A}^{-1}Y&{\tilde C}Y+{\tilde A}^{-1}X).
\end{align*}
We make a restriction by choosing $X=\text{diag}(\cos\vec\theta)$ and $Y=\text{diag}(\sin\vec\theta)$. We also note the connection to a similar decomposition given in Ref.~\cite{mehrmann1988,dopico2009} which could also be used to define a more restricted class of simulatable operations.}.
The allowed angles $\theta=(\theta_1,\dots,\theta_n)^T$ are selected from the set $\theta_j\in\Theta$ as defined in Eq.~(\ref{eq:anglesset}). Equivalently, we could restrict the symplectic matrices in Eq.~(\ref{eq:block-form}) such that the block components satisfy ${A=\tilde A\text{diag}(\cos\vec\theta)}$ and  ${B=\tilde A\text{diag}(\sin\vec\theta)}$ for some symmetric $\tilde A$. 

When including the Heisenberg-Weyl operations we denote the full set of simulatable operations as
\begin{align}
\label{eq:classB}
    \mathcal B = \text{HW}(n)\times \text{DSp}(2n,\mathbb R).
\end{align}
We will now demonstrate that when the operations $\hat U$ are selected from the class $\mathcal B$, acting on $0$-logical GKP states, followed by multimode homodyne measurement of $\hat q$, the circuit is simulatable. The class $\mathcal B$ contains operations of the form $\hat U$ in Eq.~(\ref{eq:defineU}) where $(\bar{\mathbf r},M)$ is selected from the set of possible phase space displacements $\bar{\mathbf r}\in\text{HW}(n)$ and symplectic operations given in Eq.~(\ref{eq:operations-dsp}). In Section~\ref{sec:summarysimulatable} we will show that the class $\mathcal B$ is contained in the class $\mathcal A$ defined in Eq.~(\ref{eq:classA}).

The PDF of a general state $\hat \rho$ measured in the position basis in all modes is given by
\begin{align}
    \text{PDF}(\mathbf x)
    =&\Tr(\hat \rho \ket{\hat{\mathbf{q}}=\mathbf{x}}\bra{\hat{ \mathbf{q}}=\mathbf{x}}),
\end{align}
where
\begin{align}
\label{eq:bold-q}
\ket{\hat{\mathbf{q}}=\mathbf{x}}=\prod_i\ket{\hat{q}_i=x_i}
\end{align}
and $\mathbf x=(x_1,\dots,x_n)$. Similarly to the single-mode case, rather than evolving the state we can evolve the position measurement operators. By analyzing the decomposed matrix in Eq.~(\ref{eq:operations-dsp}) we can identify two independent operations which transform the measurement quadratures in a convenient way. The action of the second matrix in Eq.~(\ref{eq:operations-dsp}) can be expressed as a series of single-mode rotations
\begin{align}
    \hat R(\vec\theta)=\hat R_1(\theta_1)\dots \hat R_n(\theta_n)
\end{align}
which rotate each mode independently according to
\begin{align}
    \label{eq:rotation-singlemode}
    \hat R^\dagger (\vec\theta)\hat q_j\hat R(\vec\theta)=\hat q_j\cos\theta_j-\hat p_j\sin\theta_j=\hat z_j^{\theta_j}.
\end{align}
We then inspect the unitary operation corresponding to the first matrix with ${\tilde A}$ in block $(1,1)$ which is given by $\hat U_{{\tilde A}}$ and has the effect
\begin{align}
    \label{eq:mode-transform}
    \hat U_{{\tilde A}}^\dagger \hat q_j\hat U_{{\tilde A}}=\sum_ia_i^{(j)}\hat q_i,
\end{align}
where $a_i^{(j)}$ is the element in the $(j,i)$ position of the matrix $\tilde A$.

Eqs.~(\ref{eq:rotation-singlemode}) and (\ref{eq:mode-transform}) allow us to express the Heisenberg measurement operators as
\begin{align}
    \hat R^\dagger (\vec\theta)\hat U_{\tilde A}^\dagger \hat q_j\hat U_{\tilde A}\hat R(\vec\theta)=\sum_ia_i^{(j)}\hat z_i^{\theta_i}.
\end{align}
We can also include arbitrary displacements in position by including a displacement operation $e^{-i\hat{\mathbf p}\cdot \mathbf c}$ before measurement, which results in the total Heisenberg measurement operator
\begin{align}
    \hat Q_j=\hat R^\dagger (\vec\theta)\hat U_{\tilde A}^\dagger e^{i\hat{\mathbf p}\cdot \mathbf c}\hat q_je^{-i\hat{\mathbf p}\cdot \mathbf c}\hat U_{\tilde A}\hat R(\vec\theta).
\end{align}
This can be expressed in terms of the individual rotated modes as
\begin{align}
    \label{eq:heisenberg-opterator-multi}
    \hat Q_j=\sum_ia_i^{(j)}\hat z_i^{\theta_i}+c_j,
\end{align}
where, as said, $a_i^{(j)}$ is the element in the $(j,i)$ position of the matrix $\tilde A$ and $\vec \theta$ are the rotation values of the rotation matrix $R(\vec \theta)$.

Writing the PDF in terms of the Heisenberg-evolved measurement operators and using the cyclic property of the trace we find
\begin{align}
    \text{PDF}(\mathbf x)=\bra{\mathbf 0_{\text{GKP}}}\ket{\hat{\mathbf Q}=\mathbf x}\bra{\hat{\mathbf Q}=\mathbf x}\ket{\mathbf 0_{\text{GKP}}},
\end{align}
where we use the same notation as Eq.~(\ref{eq:bold-q}) to denote the simultaneous eigenkets of the operators $\hat Q_j$ over $x_j$, for all $j\in\{1,\dots,n\}$.
Expressing the projection operators as Dirac delta functions, we have
\begin{align}
    \text{PDF}(\mathbf x)=\bra{\mathbf 0_{\text{GKP}}}\prod_j\delta(\hat Q_j-x_j)\ket{\mathbf 0_{\text{GKP}}}.
\end{align}
We can then substitute the expression for the Heisenberg evolved measurement operators, Eq.~(\ref{eq:heisenberg-opterator-multi}), to find
\begin{align}
    &\text{PDF}(\mathbf x)\nonumber \\
    =&\bra{\mathbf 0_{\text{GKP}}}\prod_j\delta(\sum_ia_i^{(j)}\hat z_i^{\theta_i}+c_j-x_j)\ket{\mathbf 0_{\text{GKP}}}.
\end{align}
Inserting the identity Eq.~(\ref{eq:identity-rotated-modes}) and evaluating the integral over the resulting delta functions results in the final PDF
\begin{align}
    &\text{PDF}(\mathbf x)\nonumber\\
    =&\sum_{m_1,\dots,m_n\in\mathbb Z}  \prod_{j=1}^n\delta(\sum_{i=1}^na_i^{(j)}m_i\sqrt\pi\Delta_i+c_j-x_j).
\end{align}
The full calculation details of the PDF are given in Appendix \ref{sec:appendix-multimodeoperations-pdf}. 
If the symplectic operation is given in terms of the decomposition of Eq.~(\ref{eq:operations-dsp}) then it is possible to calculate all $n$ of the parameters $\Delta_i$ in linear time by the same considerations as reported in Section \ref{sec:simulation-singlemode}. The PDF involves a $n\times n$ matrix of coefficients $a^{(j)}_{i}$ and so evaluating the PDF involves a total of $n^2$ coefficients of this form. We have therefore demonstrated an efficient method for evaluating the PDF for any circuit of the type given in Fig.~\ref{fig:multimodecircuitclass-maintext}. 

In summary, due to the form of the matrix decomposition in Eq.~(\ref{eq:operations-dsp}), we can simulate any circuit with modes initialized as $0$-logical GKP states acted upon by parallel single-mode rotations parameterized by $\theta_j\in\Theta$, followed by any symplectic operation parametrized by a symplectic matrix which when written in block form, as in Eq.~(\ref{eq:block-form}), has block component $B=0$. The restriction to generate a symplectic matrix such that the block component $B=0$ is equivalent to ensuring that the operation does not transform $\hat q_j\rightarrow\hat p_k$ or vice versa, for any $j,k$. The SUM gate satisfies these criteria. Therefore, we can construct arbitrary examples of circuits that are restricted to the form in Eq.~(\ref{eq:operations-dsp}) by first choosing single-mode rotations parameterized by $\theta_j\in\Theta$, followed by any combination of squeezing operations and SUM gates. An example simulation of a Gaussian operation acting on GKP states is given in Appendix  \ref{sec:appendix-multimodeoperations-example}.   
A circuit diagram of a general circuit chosen from the set of allowed operations $\mathcal B$ is given in Appendix \ref{sec:appendix-circuit-diagrams-B}.

Similar to the result given in Ref.~\cite{bartlett2002}, it is clear that certain feed-forward operations can be efficiently simulated too, when they can be substituted with Gaussian two-mode operations followed by deferred measurements. In particular, Pauli operators implemented after classical feed-forward can still be simulated efficiently. 

It may also be possible to provide an alternative proof of the simulatability of these circuits by adapting the formalism developed in Ref.~\cite{Juani-thesis} to input GKP states. 

\section{Summary of simulatable operations}
\label{sec:summarysimulatable}
We have defined two new restricted classes of Gaussian operations, given in Eq.~(\ref{eq:classA}) and (\ref{eq:classB}), in terms of sets of restricted classes of symplectic operations, $\text{RSp}(2n,\mathbb R)$ and $\text{DSp}(2n,\mathbb R)$.
Operations selected from the restricted class $\mathcal A$, applied to $0$-logical GKP states, followed by measurement of a single mode are simulatable.  Operations selected from the restricted class $\mathcal B$, applied to $0$-logical GKP states, followed by measurement of multiple modes are simulatable.

In Ref.~\cite{garcia-alvarez2020} it is proven that using the Gottesman-Knill theorem, one can simulate any continuous-variable quantum computation initialized to $0$-logical GKP eigenstates, acted upon by encoded Clifford group operations $\mathcal C_d$ for GKP-encoded qudits in arbitrary dimension $d$, and measured with homodyne measurement.

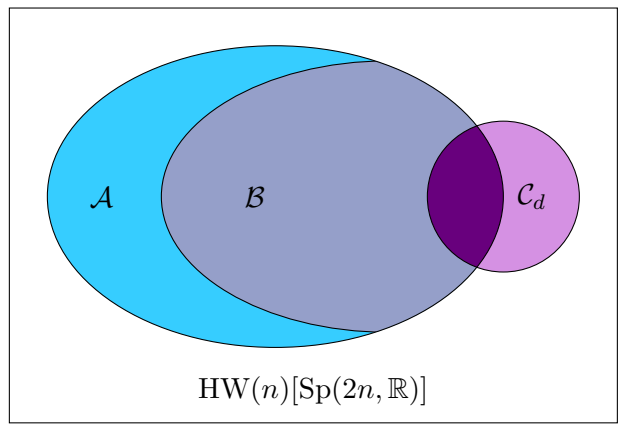
\begin{figure}[h]
    \centering
    \begin{tikzpicture}[fill=gray]
    \draw (-3.5,-3) rectangle (4.5,2.5);
    \fill[fill=blueA] (0,0) ellipse (3 and 2);
    \scope
    \clip (0,0) ellipse (3 and 2);
    \fill[fill=blueB] (1.5,0) ellipse (3 and 1.8);
    \endscope
    \fill[fill=blueC] (3,0) circle (1);
    \scope
    \clip (3,0) circle (1);
    \fill[fill=blueCB] (0,0) ellipse (3 and 2);
    \endscope
    \node[below] at (0.5,-2.25) {$\text{HW}(n)[\text{Sp}(2n,\mathbb R)]$};
    
    \draw (0,0) ellipse (3 and 2);
    \scope
    \clip (0,0) ellipse (3 and 2);
    \draw (1.5,0) ellipse (3 and 1.8);
    \endscope
    \draw (3,0) circle (1);
    
    \node[left] at (-2,0) {$\mathcal A$};
    \node[left] at (0,0) {$\mathcal B$};
    \node[left] at (3.7,0) {$\mathcal C_d$};
    \end{tikzpicture}
    \caption{The classes of circuits we consider are displayed as a Venn diagram. We know that all $\mathcal A,\mathcal B,\mathcal C_d$ are contained within the set of Gaussian operations $\text{HW}(n)[\text{Sp}(2n,\mathbb R)]$. We also know that the logical encoded Clifford group does not completely contain, nor is completely contained by $\mathcal A$ or $\mathcal B$. Finally, we know that $\mathcal B$ is contained by $\mathcal A$. Note that the size of each of these regions in the diagram is arbitrary. It is however meant to represent the fact that the parameters of the symplectic and displacement operations are selected respectively from a zero-measure set and the set of rational numbers, for $\mathcal C_d$, and from a dense set and all reals for $\mathcal A$.
    }
    \label{fig:venn}
\end{figure}
The relationships between the classes $\mathcal A,\mathcal B,\mathcal C_d$ demonstrate the power of the new simulation techniques outlined in this paper. First, we know that all $\mathcal A$ and $\mathcal B$ and $\mathcal{C}_d$ are contained within the full set of Gaussian operations
\begin{align}
    \mathcal A,\mathcal B,\mathcal{C}_d &\subset \text{HW}(n)[\text{Sp}(2n,\mathbb R)].
\end{align}
We also note that
\begin{align}
    \mathcal B \subset \mathcal A& \quad \text{and}\label{eq:BA}\\
    \mathcal A,\mathcal B \not\subset \mathcal C_d&.\label{eq:CA}
\end{align}
The relationship of Eq.~(\ref{eq:BA}) can be understood by considering that any symplectic matrix in $\text{DSp}(2n,\mathbb R)$, which satisfies the constraints of Eq.~(\ref{eq:operations-dsp}), must also satisfy the constraints of $\text{RSp}(2n,\mathbb R)$, given in Eq.~(\ref{eq:rspconstraints-maintext}). This is because the constraints of $\text{DSp}(2n,\mathbb R)$ given for the first row of the symplectic matrix are equivalent to all of the constraints of $\text{RSp}(2n,\mathbb R)$. Furthermore, the relationship of Eq.~(\ref{eq:CA}) can be seen immediately by considering that, for example, arbitrary displacements, which are contained in $\mathcal A$ and $\mathcal B$, are not contained within $\mathcal C_d$.

These relationships are presented as a Venn diagram in Fig.~\ref{fig:venn}.
Eq.~(\ref{eq:CA}) shows that there exists a large class of circuits which previously were not known to be simulatable which are simulatable by the methods detailed in this paper.

We will now compare the results of this work with those of Ref.~\cite{garcia-alvarez2020}. It is always possible to define Gaussian operations in the form of Eq.~(\ref{eq:defineU}) in terms of the real parameters of the symplectic matrix and displacements. The methods described in Ref.~\cite{garcia-alvarez2020} allow us to simulate only those parameters selected from a zero-measure set for the symplectic matrices and all rational displacements. In contrast, the methods described in this work demonstrate that all circuits selected from $\mathcal A$ followed by single-mode measurement are simulatable. The parameters of $\mathcal A$ constitute a set which is dense on the reals and thus significantly expands the class of simulatable Gaussian operations. The proof of this fact is given in Appendix \ref{sec:appendix-restrictedsets-density}. Note, however, that the density of the parameters defining the operators should be considered a mathematical property characterizing the class of simulatable operations. The density does not imply that it is possible to simulate the PDF of operations outside of the set, i.e. it does not mean that operations in the closure of the set are necessarily simulatable.

Furthermore, in Appendix \ref{sec:appendix-restrcitedsets-clifford} we show that
\begin{align}
    \mathcal C_d \not\subset \mathcal A\quad  \text{and}\quad \mathcal C_d \not\subset \mathcal B
\end{align}
which informs us that neither $\mathcal A$ nor $\mathcal B$ completely contain the Clifford group operations which are simulatable by previous methods. 

The classes described in this section are those that we have shown to be simulatable in our work and are not necessarily fundamental. It is possible that there exists a new class of simulatable continuous-variable operations acting on $0$-logical GKP states, e.g. $\mathcal D$, which contains all of $\mathcal A,\mathcal B,\mathcal C_d$, which remains undiscovered.

\section{Realistic GKP states} \label{sec:realistic-gkp}

Here we briefly sketch how to extend the analysis that we have performed to the case of realistic GKP states. In principle, our simulation method could be extended to such case.
We show that an analytical expression for the PDF of a rotated and squeezed
single-mode GKP state, analogous to Eq.~(\ref{eq:singlemodepdf-main}), can also be derived for the case of realistic GKP states. However, due to the more complex form of this PDF, which displays Gaussian peaks rather than a Dirac comb, we cannot directly apply the same derivation as before to evaluate the PDF after multiple modes have undergone symplectic operations.

The rotated and squeezed wave function of a realistic GKP state is given by
\begin{align}
    \psi_{\theta,s}(x)&=\bra{q=x}\hat S(s)
    \hat R(\theta)\ket{ 0_{\text{GKP}}^\Delta}
\end{align}
which can --- as in the case of infinitely squeezed GKP states, see Eq.~(\ref{eq:justtheta}) --- be expressed in terms of the wave function $\psi_\theta^\Delta(x)$ of the rotated GKP state, i.e.
\begin{align}
    \psi_{\theta,s}^\Delta(x)=\psi_\theta^\Delta(x/s).
\end{align}
In Appendix \ref{appendix:route-to-realistic} we demonstrate that it is possible to write this expression analytically in the form
\begin{align}
    \psi^\Delta_{\theta}(x)=e^{x^2\gamma}\vartheta(\zeta=x\eta,\tau)
\end{align}
which is defined in terms of the constants
\begin{align}
    \eta=&-\csc\theta/\left(\sqrt\pi (s+\Delta^4 s-i\Delta^2 s\cot\theta)\right),\nonumber\\
        \tau=&2i(\Delta^2-i\cot\theta)/(1+\Delta^4-i\Delta^2\cot\theta), \nonumber\\
        \gamma=&\frac{i(i\Delta^2 +(1+\Delta^4)\cot\theta)}{2 s^2(1+\Delta^4-i\Delta^2\cot\theta)}
\end{align}
and results in a single-mode rotated realistic GKP state having a PDF of the form
\begin{align}
    |\psi^\Delta_{\theta}(x)|^2
    &\propto e^{x^2(\gamma+\gamma^*)}|\vartheta(\zeta=x\eta,\tau)|^2.
\end{align}

The PDF of a circuit with input realistic GKP states followed by Gaussian operations and homodyne measurements is given by
\begin{align}
    \text{PDF}(\mathbf x)=\bra{\mathbf 0_{\text{GKP}}^\Delta}\prod_j\delta(\hat Q_j-x_j)\ket{\mathbf 0_{\text{GKP}}^\Delta}.
\end{align}
Solving this expression analytically would require solving integrals over a product of Jacobi theta functions, which hinders us from using the same techniques used for the ideal case. 
We are unaware of a method to evaluate this expression. We provide a more detailed overview of the challenges of evaluating the PDF for realistic GKP states, for both single-mode and multimode measurement, in Appendix \ref{appendix:route-to-realistic}.\\

\section{Conclusion}
\label{sec:conclusion}

In this work, we have shown that large classes of Gaussian operations, in combination with input $0$-logical GKP states and homodyne measurements, are classically efficiently simulatable. The proof structure used to prove our result is based on decompositions of the symplectic matrix and directly calculating the corresponding PDF, hence introducing very different methods compared to the ones used to prove the simulatability of a restricted class of Gaussian circuits in Ref.~\cite{garcia-alvarez2020}. As a matter of fact, compared to ${\mathcal C}_d$, for single-mode measurements we define a larger class $\mathcal A$ of simulatable operations which includes symplectic operations with parameters which are dense in the reals and all continuous displacements. Meanwhile, for multimode measurements, we define a new set $\mathcal B$, not contained by ${\mathcal C}_d$ which is also simulatable. 

Identifying circuits which are simulatable is, beyond fundamentally relevant, also useful for the development of quantum computers, providing a possibility for benchmarking and verification. Indeed a quantum computer can  be initially programmed to solve problems which are simulatable with classical computers, as a test to identify whether the computation result is accurate~\cite{arute2019}.

Our proof of efficient simulatability only holds for the case of infinitely squeezed $0$-logical GKP states, and we have sketched the difficulties encountered when trying to extend this result to the case of realistic, i.e. finitely squeezed, GKP states~\cite{gottesman2001}. Establishing conclusively whether realistic GKP states  are also simulatable following Gaussian operations is hence left for future work.

We acknowledge useful discussions with Laura García-Álvarez and Juani Bermejo-Vega.
G. F. and C. C. acknowledge support from the VR (Swedish Research Council) Grant QuACVA. G. F. and C. C. acknowledge support from the Wallenberg Center for Quantum Technology (WACQT). 

\bibliographystyle{quantum}
\bibliography{./main.bib}


\appendix

\begin{widetext}
\renewcommand\thesubsubsection{\roman{subsubsection}}
\section{Evaluating the PDF of a rotated and squeezed GKP state}
\label{sec:appendix-singlemodeops}
\subsection{Transforming a single mode}
\label{sec:appendix-singlemodeops-transforming}
    We begin by analyzing a simplified version of the circuit in Fig.~\ref{fig:singlemodecircuitclass-maintext} where we have one mode and $\hat U$ is the unitary operator corresponding to a single-mode symplectic operation. This would correspond to a circuit of the form given in Fig.~\ref{fig:circuitsinglemode}. We will then generalize to include displacement in phase space (i.e. displacement in $\hat q$ or $\hat p$).
    
    The general form of the transformed quadrature under symplectic operations for an individual mode is given by
    \begin{align}
        \hat U^\dagger \hat q \hat U =a \hat q +b \hat p. \label{eq:singlemode}
    \end{align}
    If we are given or can calculate the real coefficients $a,b\in\mathbb R$ in Eq.~(\ref{eq:singlemode}), then it is possible to reconstruct a single-mode operation $\hat U$ which could generate such a transformation. By using only a certain set of operations we can calculate a decomposition of $\hat U$ which will help us to calculate the PDF after measurement of the position quadrature by applying that operation to the GKP state and calculating its wave function and hence also its PDF.
    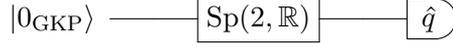
\begin{figure}[h!]
     \centering
     $$
     \Qcircuit @C=1.5em @R=1.2em {
     \lstick{\ket{0_{\text{GKP}}}} & \qw & \gate { \text{Sp}(2,\mathbb R)} & \qw & \measureD{\hat q}
     }
     $$
     \caption{The measurement of a single mode following an arbitrary symplectic operation acting on the $\ket{0_{\text{GKP}}}$ state can be tracked by considering the evolution of the position quadrature in the Heisenberg representation.}
     \label{fig:circuitsinglemode}
    \end{figure}
    
    We will begin by first deriving an appropriate decomposition, inspired by Ref.~\cite{arvind1995} (but ignoring the shear gate which acts solely on the $\hat p$ quadrature), which can be applied to the symplectic operation for all values of $a,b\in\mathbb R$.
    
    Within the symplectic formalism, the rotation operator $\hat R(\theta)=e^{i\theta(\hat q^2+\hat p^2)/2}$ has the effect of transforming the mode quadratures as
    \begin{align}
        \mqty(\hat q'\\\hat p')=\mqty(\cos\theta & -\sin\theta\\\sin\theta & \cos\theta)\mqty(\hat q \\ \hat p),
    \end{align}
    while the squeezing operator $\hat S(s)=e^{-i\frac{\ln s}{2}(\hat q\hat p+\hat p\hat q)}$ has the effect of
    \begin{align}
        \mqty(\hat q'\\\hat p')=\mqty(s & 0 \\ 0 & 1/s)\mqty(\hat q \\ \hat p),
    \end{align}
    meaning that the operation $\hat S(s)\hat R(\theta)$ has the effect
    \begin{align}
        \mqty(\hat q'\\\hat p')=\mqty(s & 0 \\ 0 & 1/s)\mqty(\cos\theta & -\sin\theta\\\sin\theta & \cos\theta)\mqty(\hat q \\ \hat p)=\mqty(s & 0 \\ 0 & 1/s)\mqty(\hat q \cos\theta-\hat p \sin\theta \\ \hat q \sin\theta +\hat p \cos\theta)=\mqty(\hat q s\cos\theta-\hat p s\sin\theta \\ \frac 1 s\left(\hat q \sin\theta +\hat p \cos\theta\right)).
    \end{align}
    This means that we can obtain the evolved quadrature $\hat Q$ by acting on $\hat q$ with  a rotation and squeezing operation
    \begin{align}
        \hat Q=\hat R^\dagger (\theta)\hat S^\dagger(s)\hat q \hat S(s)\hat R(\theta)=\hat q s\cos\theta-\hat p s\sin\theta.
    \end{align}
    This allows us to parameterize the operator $\hat U=\hat S(s)\hat R(\theta)$ where 
    \begin{align}
    \label{eq:a-b-vs-theta-s}
    a=s\cos\theta; \nonumber \\
    b=-s\sin\theta.
     \end{align}
    Then, we can calculate the wave function of the transformed mode as
    \begin{align}
        \psi_{\theta,s}(x)&=\bra{q=x}\hat S(s)
        \hat R(\theta)\ket{0_{\text{GKP}}}\nonumber\\
        &=\bra{q=x/s}\hat R(\theta)\ket{0_{\text{GKP}}}\nonumber\\
        &=\bra{q=x/s}\hat R(\theta)\sum_m\ket{q=2m\sqrt\pi}.
    \end{align}
    We can interpret calculating each summand as a wave mechanics problem with Hamiltonian ${\hat H=\frac 1 2(\hat q^2+\hat p^2)}$ parameterized with $\theta$. This becomes
    \begin{align}
        \bra{q=x/s}\hat R(\theta)\ket{q=x'}=
        \bra{q=x/s}e^{-i\hat H\theta}\ket{q=x'}=K(x/s,x';\theta)
    \end{align}
    which is the propagator for the simple harmonic oscillator~\cite{sakurai} and can be evaluated to
    \begin{align}
        \label{eq:propogator}
        K(x,x';\theta)=\frac{1}{\sqrt{2\pi i \sin\theta}}\exp(\frac{i}{2\sin\theta}\left((x^2+x'^2)\cos\theta-2xx'\right)).
    \end{align}
    By completing the square, we can rewrite the propagator in the form
    \begin{align}
        K(x,x';\theta)=&\frac{1}{\sqrt{2\pi i \sin\theta}}\exp(\frac{i\cot\theta}{2}\left(x^2+x'^2-2xx'\sec\theta\right))\nonumber\\
        =&\frac{1}{\sqrt{2\pi i \sin\theta}}\exp(\frac{i\cot\theta}{2}\left((x-x'\sec\theta)^2+(1-\sec^2\theta)x'^2\right))\nonumber\\
        =&\frac{1}{\sqrt{2\pi i \sin\theta}}\exp(\frac{i\cot\theta}{2}\left((x-x'\sec\theta)^2-\tan^2\theta x'^2\right))\nonumber\\
        =&\frac{1}{\sqrt{2\pi i \sin\theta}}\exp(\frac{i}{2}\left((x-x'\sec\theta)^2\cot\theta-\tan\theta x'^2\right))
    \end{align}
    which gives the expression for the wave function

    \begin{align}
        \label{eq:rotatedwf}
        \psi_{\theta,s}(x)=&\frac{1}{\sqrt{2\pi i\sin\theta}}\sum_{m} \exp(\frac{i}{2}\left((x/s-2m\sqrt\pi\sec\theta)^2\cot\theta-4\pi m^2\tan\theta \right))\nonumber\\
        =&\frac{1}{\sqrt{2\pi i\sin\theta}}\sum_{m} \exp(\frac{i}{2}\left(x^2\cot\theta/s^2-4x m\sqrt\pi\csc\theta/s+4m^2\pi(\csc\theta\sec\theta-\tan\theta) \right))\nonumber\\
        =&\frac{1}{\sqrt{2\pi i\sin\theta}}\sum_{m} \exp(\frac{i}{2}\left(x^2\cot\theta/s^2-4x m\sqrt\pi\csc\theta/s+4m^2\pi\cot\theta \right))\nonumber\\
        =&\frac{1}{\sqrt{2\pi i\sin\theta}}e^{ix^2\cot\theta/(2s^2)}\sum_{m} e^{\pi im^2 2\cot\theta }e^{-2\pi imx \csc\theta/(s\sqrt\pi)}\nonumber\\
        =&\frac{1}{\sqrt{2\pi i\sin\theta}}e^{ix^2\cot\theta/(2s^2)} \vartheta(\zeta=- x \csc\theta/(s\sqrt\pi);\tau=2\cot\theta)
    \end{align}
    where we have used the definition of the theta function,i.e.
    \begin{align}
        \label{eq:thetafunc}
        \vartheta(\zeta;\tau)=\sum_{m\in \mathbb Z}e^{\pi i m^2 \tau}e^{2\pi i m \zeta }.
    \end{align}

    This gives a succinct analytic expression for the PDF of measuring the $\hat q$ quadrature of a GKP state under any arbitrary single-mode symplectic operation,
    \begin{align}
    \label{eq:probtheta-original}
        |\psi(x)|^2
        =&\frac{1}{4\pi \sin\theta} \abs{\vartheta(\zeta=-x \csc\theta /(s\sqrt\pi);\tau=2\cot\theta)}^2.
    \end{align}
    In order to incorporate phase space displacements we note that any Gaussian operation involving a symplectic operation
    \begin{align}
        \hat U_M\hat D_{\mathbf r}
    \end{align}
    can be reordered while preserving the structure of the symplectic operation~\cite{serafini2017}, i.e.
    \begin{align}
        \label{eq:commutation-displacements}
        \hat U_M\hat D_{{\mathbf r}}=\hat D_{M^{-1}{\mathbf r}}\hat U_M.
    \end{align}
    Therefore, any Gaussian operation acting on a single mode, for which we will measure the $\hat q$ quadrature, can be decomposed as
    \begin{align}
        \hat U=\hat D_{\bar{\mathbf r}}\hat S(s)\hat R(\theta).
    \end{align}
    The PDF of measuring the $\hat q$ quadrature of a single-mode GKP state under any Gaussian operation can therefore be written as
    \begin{align}
    \label{eq:probtheta}
        |\psi(x)|^2
        =&\frac{1}{4\pi \sin\theta} \abs{\vartheta(\zeta=- (x-c) \csc\theta /(s\sqrt\pi);\tau=2\cot\theta)}^2,
    \end{align}
    where $\bar r_1=c$ is the magnitude of displacement in position.
    
    The PDF involving the theta function in Eq.~(\ref{eq:probtheta}) can be evaluated for certain values of the angle $\theta$, or equivalently certain values of $\tau$. The angle
    $\theta$ cannot take values of the form $k\pi$ for integer $k\in\mathbb Z$, but as we shall see later, these rotations are trivial to evaluate using a different method.
    
    We now consider different cases where we can evaluate the PDF. 
    
    \textbf{(Case 1:} $\cot\theta=u/v$ s.t. $u\in\mathbb Z, v\in\mathbb Z_{\text{odd}}$ \textbf{)}
    The theta function can be reduced to a sum over Gauss sums whenever $\cot\theta=u/v$ is a fraction, i.e. a rational number.
    We also assume that the fraction $u/v$ is written in its simplest form, i.e. $\gcd(u,v)=1$. If it is not then it should be reduced as such by first calculating $\gcd(u,v)$, which can be calculated in polynomial time with respect to the size of $u,v$~\cite{arora2009}. Then, $\exp(\pi i \tau)=\exp(2\pi i u/v )$ becomes a root of unity and so the summation repeats in cycles~\cite{stackexchangejacobi}. This can be utilized by expanding the summation over $n$ into an infinite summation over $n'$ and a summation over $m\in \mathbb Z_{ v}=\{0,1,\dots, v-1\}$
    \begin{align}
    \label{eq:thetaandgauss}
        \vartheta(\zeta;\tau=2u/v)=&\sum_{n\in \mathbb Z}e^{\pi i n^2 \tau}e^{2\pi i n \zeta }\nonumber\\
        =&\sum_{n\in \mathbb Z}e^{2\pi i n^2 u/v}e^{2\pi i n \zeta }\nonumber\\
        =&\sum_{n'}\sum_{m\in \mathbb Z_{v}} e^{2\pi i (n'v+m)^2 u/v}e^{2\pi i (vn'+m) \zeta}\nonumber\\
        =&\sum_{n'}\sum_{m\in \mathbb Z_{v}} e^{2\pi i (v^2n'^2+2mvn'+m^2) u/v}e^{2\pi i (vn'+m) \zeta}\nonumber\\
        =&\sum_{n'}\sum_{m\in \mathbb Z_{v}} e^{2\pi i (vn'^2u+2mn'u+m^2u/v)}e^{2\pi i (vn'+m) \zeta}\nonumber\\
        =&\sum_{n'}\sum_{m\in \mathbb Z_{v}} e^{2\pi im^2u/v}e^{2\pi i vn' \zeta}e^{2\pi i m\zeta}\nonumber\\
        =&\sum_{m\in \mathbb Z_{v}} e^{2\pi im^2u/v}e^{2\pi i m\zeta}\sum_{n'}e^{2\pi i vn' \zeta}\nonumber\\
        =&\sum_{m\in \mathbb Z_{v}} e^{2\pi im^2u/v}e^{2\pi i m\zeta}\sum_{n'}\delta(\zeta-n'/v)\nonumber\\
        =&G(\zeta,\tau)\sum_{n'} \delta(\zeta-n'/v),
    \end{align}
    where we have identified the Gauss sum~\cite{apostol1986,berndt1998,iwaniec2004}
    \begin{align}
        G(\zeta=x\sec\theta/(2s\sqrt\pi), \tau=2u/v)=\sum_{m\in \mathbb Z_{v}} \exp(2\pi i m^2 u/v)\exp(2\pi i m \zeta).
    \end{align}
    
    By using the fact that the delta function evaluates to $0$ for all values of $\zeta$ not equal to $n'/v$ for some $n'\in \mathbb Z$, we can rewrite the theta function as
    \begin{align}
    \label{eq:thetaandgaussonly}
        \vartheta(\zeta;\tau=2u/v)
        =&\sum_{n'}G(\zeta=n'/v,\tau) \delta(\zeta-n'/v),
    \end{align}
    where the Gauss sum needs only to be identified for certain values of $\zeta$, rather than the entire real axis. For these values of $\zeta$, the Gauss sum is given by
    \begin{align}
        G(\zeta=n'/v,\tau=2u/v)=\sum_{m\in \mathbb Z_{v}} \exp(2\pi i m^2 u/v)\exp(2\pi i m n'/v).
    \end{align}
    Since we have confined to the case that $\gcd(u,v)=1$ and $v\in\mathbb Z_{\text{odd}}$ then we immediately have the result that
    \begin{align}
        |G(\zeta=n'/v,\tau=2u/v)|=\sqrt v \abs{\left(\frac{u}{v}\right)}>0 \quad \text{ for } \gcd(u,v)=1, v\in\mathbb Z_{\text{odd}}\text{ and }n',u\in\mathbb Z
    \end{align} is a fixed constant dependent only on $u,v$~\cite{berndt1998}, where $\left(\frac{u}{v}\right)$ denotes the Jacobi symbol, a generalization of the Legendre symbol.
    
    Then, we can substitute this into Eq.~(\ref{eq:thetaandgaussonly})
    \begin{align}
        |\vartheta(\zeta;\tau=2u/v)|=&\sum_{n'} \delta(\zeta-n'/v) |G(\zeta=n'/v,\tau) |\nonumber\\
        \propto&\sum_{n'} \delta(\zeta-n'/v).
    \end{align}
    
    We can reinsert the theta function into the probability in Eq.~(\ref{eq:probtheta}) to get
    \begin{align}
        |\psi_{\theta,s}(x)|^2
        \propto& \abs{\vartheta(\zeta;\tau=2\cot\theta)}^2\nonumber\\
        \propto&\sum_{m} \delta(- x \csc\theta /(s\sqrt\pi)-m/v)\nonumber\\
        \propto&\sum_{m} \delta( x -sm\sqrt\pi\sin\theta/v).
    \end{align}
    
    \textbf{(Case 2: $\theta=0,\pi$)}
    When $\theta=0,\pi$ the probability function appears to break down because $1/\sin\theta$ is undefined. However, we can clearly identify that $\theta=0$ is simply the identity. For $\theta=\pi$, this is equivalent to applying the Fourier transform twice, which on the $\ket{0_{\text{GKP}}}$ state has the effect of the identity. In both cases we obtain
    \begin{align}
        \abs{\psi_{\theta,s}(x)}^2
        =&\abs{\bra{q=x}S(s)R(\theta=\pi)\ket{0_{\text{GKP}}}}^2\nonumber\\
        =&\abs{\bra{q=x}S(s)\ket{0_{\text{GKP}}}}^2\nonumber\\
        =&\abs{\bra{q=x/s}\ket{0_{\text{GKP}}}}^2\nonumber\\
        \propto &\sum_m\delta(x-2m\sqrt\pi s).
    \end{align}
    
    \subsection{Evaluation of the separation between the peaks}
\label{sec:appendix-singlemodeops-separation}
    We can now encompass both cases above, and write a general PDF for all rotation angles $\theta\in\Theta$ that we define in Eq.~(\ref{eq:anglesset}), in terms of the separation between peaks. Note that the squeezing parameter corresponds to a rescaling of the wave function in position space, and we can trivially write the squeezed and rotated GKP PDF in terms of the rotated PDF, i.e.
    \begin{align}
        \abs{\psi_{\theta,s}(x)}^2\propto&\abs{\psi_{\theta}(x/s)}^2.
    \end{align}
    
    Ignoring the normalization constant, as in~\cite{gottesman2001}, we can write a formula for the PDF of a general rotated mode
    \begin{align}
    \label{eq:finalprobsinglemode}
        \abs{\psi_{\theta}(x)}^2=&\sum_{m} \delta(x-m\sqrt\pi \Delta)
    \end{align}
    with
    \begin{align}
    \label{eq:deltacases}
        \Delta=\begin{cases}
        \sin(\theta) /v\quad & \text{ (case 1) if } \cot\theta=u/v : u\in\mathbb Z,v\in\mathbb Z_{\text{odd}} \text{ and } \gcd(u,v)=1,\\
        2\quad &\text{ (case 2) if } \theta=k\pi \text{ for } k \in \mathbb Z.
        \end{cases}
    \end{align}
    
    Note that in the special case when $\theta=\pi/2$ we have $\cot(\pi/2)=0$, so we identify $u=0$ and $v=1$, as the simplest form of the fraction $u/v$, to arrive at $\Delta=1/s$. This coincides with what we would expect when performing a Fourier transform (equivalent to a $\pi/2$ rotation) on the input state in the logical basis $F\ket{0_{\text{GKP}}}=\ket{+_{\text{GKP}}}=\sum_m \ket{q=m\sqrt\pi}$ followed by a measurement.

\section{Simulation of measurement of a single mode for Gaussian operations in $\mathcal A$}
\label{sec:appendix-singlemodemeasurement}
    
    \subsection{Calculating the PDF}
    \label{sec:appendix-singlemode-pdf}
    
    
    For the class of circuits that we consider in Section \ref{sec:simulation-singlemode}, namely those belonging to the class $\mathcal A$ defined in the main text, the PDF is given by
    \begin{align}
        \text{PDF}(x_1)=&\Tr(\hat \rho \ket{\hat q_1=x_1}\bra{\hat q_1=x_1}),
    \end{align}
    where $\hat \rho$ is the state of all of the modes after the multimode operation $\hat U$ (i.e. ${\hat \rho=\hat U\hat \rho_0\hat U^\dagger}$, where $\hat \rho_0$ is the density matrix of $n$ modes prepared in the $0$-logical GKP state $\ket{0_{\text{GKP}}}$), and $\ket{\hat q_1=x_1}$ is the position eigenket for the first mode. Using the cyclic property of the trace, and then evaluating it as a trace over all the other modes in the position basis, we find that the PDF can be written as
     \begin{align}
    \label{eq:probability-density-multimode}
        \text{PDF}(x_1)=&\Tr(\hat U\hat\rho_0\hat U^\dagger \ket{\hat q_1=x_1}\bra{\hat q_1=x_1})\nonumber \\
        =&\Tr(\bra{\hat q_1=x_1}\hat U\hat \rho_0\hat U^\dagger \ket{\hat q_1=x_1})\nonumber \\
        =&\int \dd x_2 \dots \int \dd x_{n} \bra{\hat q_1=x_1}\dots\bra{\hat q_n=x_n}\hat U\hat \rho_0\hat U^\dagger \ket{\hat q_1=x_1}\dots \ket{\hat q_n=x_n}.
    \end{align}
    
    The PDF can then be evaluated by inserting the identity over all rotated quadratures $\hat z^{\theta_i}_i$ which is given by
    \begin{align}
        \mathbbm 1 = \prod_{j=1}^n \int \dd z_j \ket{\hat z_j^{\theta_j}=z_j}\bra{\hat z_j^{\theta_j}=z_j}=  \int \dd \mathbf z\prod_{j=1}^n \ket{\hat z_j^{\theta_j}=z_j}\bra{\hat z_j^{\theta_j}=z_j},
    \end{align}
    and by writing the projection operators as a Dirac delta function~\cite{ferraro2005},
    \begin{align}
        \ket{\hat Q_j=x_j}\bra{\hat Q_j=x_j}=\delta\left(\hat Q_j-x_j\right).
    \end{align}
    Using the expression for the Heisenberg evolved operator Eq.~(\ref{eq:heisenberg-operator-single}) allows us to evaluate the PDF as
    \begin{align}
        &\text{PDF}(\hat Q_1=x_1)\nonumber \\
        =& \int \dd\mathbf z\bra{0_{\text{GKP}}}\dots \bra{0_{\text{GKP}}}\ket{\hat Q_1=x_1}\bra{\hat Q_1=x_1}\prod_{j=1}^n \ket{\hat z_j^{\theta_j}=z_j}\bra{\hat z_j^{\theta_j}=z_j}\ket{0_{\text{GKP}}}\dots \ket{0_{\text{GKP}}}\nonumber\\
        =& \int \dd\mathbf z\bra{0_{\text{GKP}}}\dots \bra{0_{\text{GKP}}}\delta(\hat Q_1-x_1)\prod_{j=1}^n \ket{\hat z_j^{\theta_j}=z_j}\bra{\hat z_j^{\theta_j}=z_j}\ket{0_{\text{GKP}}}\dots \ket{0_{\text{GKP}}}\nonumber\\
        =& \int \dd\mathbf z\bra{0_{\text{GKP}}}\dots \bra{0_{\text{GKP}}}\delta\left(\sum_{i=1}^{n}s_i\hat z^{\theta_i}_i+c-x_1\right)\prod_{j=1}^n \ket{\hat z_j^{\theta_j}=z_j}\bra{\hat z_j^{\theta_j}=z_j}\ket{0_{\text{GKP}}}\dots \ket{0_{\text{GKP}}}\nonumber\\
        =& \int \dd\mathbf z\bra{0_{\text{GKP}}}\dots \bra{0_{\text{GKP}}}\delta\left(\sum_{i=1}^{n}s_i z_i+c-x_1\right)\prod_{j=1}^n \ket{\hat z_j^{\theta_j}=z_j}\bra{\hat z_j^{\theta_j}=z_j}\ket{0_{\text{GKP}}}\dots \ket{0_{\text{GKP}}}\nonumber\\
        =& \int \dd\mathbf z \delta\left(\sum_{i=1}^{n}s_i z_i+c-x_1\right)\prod_{j=1}^n\abs{\bra{0_{\text{GKP}}} \ket{\hat z_j^{\theta_j}=z_j}}^2.
        \label{eq:pdf-heisenbergpic-single}
    \end{align}
    Then inserting the PDF of the single-mode rotated 0-logical GKP state given in Eq.~(\ref{eq:finalprobsinglemode}), we can express the PDF of the single-mode measurement on a multimode circuit as
    \begin{align}
        \text{PDF}(\hat Q_1=x_1)
        =& \int \dd\mathbf z \delta\left(\sum_{i=1}^{n} s_iz_i+c-x_1\right)\prod_{j=1}^n
        \sum_{m_j\in\mathbb Z} \delta(z_j-m_j\sqrt\pi \Delta_j)\nonumber\\
        =& \int \dd\mathbf z \delta\left(\sum_{i=1}^{n} s_iz_i+c-x_1\right)
        \sum_{m_1\in\mathbb Z} \delta(z_1-m_1\sqrt\pi \Delta_1)\dots \sum_{m_n\in\mathbb Z} \delta(z_n-m_n\sqrt\pi \Delta_n)\nonumber\\
        =&
        \sum_{m_1,\dots,m_n\in\mathbb Z}\delta\left(\sum_{i=1}^{n} s_im_i\sqrt\pi \Delta_i+c-x_1\right)
    \end{align}
    where we have the $\Delta_j$ specified by Eq.~(\ref{eq:deltacases}) for each mode, i.e.
    \begin{align}
    \label{eq:deltacasesrestricted}
        \Delta_j=\begin{cases}
        \sin(\theta_j) /v_j\quad & \text{ (case 1) if } \cot\theta_j=u_j/v_j : u\in\mathbb Z,v_j\in\mathbb Z_{\text{odd}} \text{ and } \gcd(u_j,v_j)=1,\nonumber\\
        2\quad &\text{ (case 2) if } \theta_j=k\pi \text{ for } k \in \mathbb Z.
        \end{cases}
    \end{align}
    
    \subsection{Simple Example}
    \label{sec:appendix-singlemode-example}
    In this section, we provide a simple example to illustrate how our decomposition methods works.
    We would like to simulate the following operations
    \begin{align}
        \hat U=e^{-i\hat q_2\hat p_1}\hat F_2
    \end{align}
    acting on GKP input states, i.e.
    \begin{align}
        \hat U\ket{0_{\text{GKP}}}\ket{0_{\text{GKP}}}.
    \end{align}
    For simplicity in this example case, these operations are chosen to be Clifford operations in terms of the encoded qubit. By considering the logical operations only, we know that the result of this circuit should be
    \begin{align}
        e^{-i\hat q_2\hat p_1}\hat F_2\ket{0_{\text{GKP}}}\ket{0_{\text{GKP}}}= e^{-i\hat q_2\hat p_1}\ket{0_{\text{GKP}}}\ket{+_{\text{GKP}}}=\frac{1}{\sqrt 2}\left(\ket{0_{\text{GKP}}}\ket{0_{\text{GKP}}}+\ket{1_{\text{GKP}}}\ket{1_{\text{GKP}}}\right),
    \end{align}
    so that measurement of the quadrature $\hat q$ on a single mode (e.g. the first mode) will give a PDF
    \begin{align}
        \text{PDF}(x_1)=\sum_m \delta(x_1-m\sqrt \pi).
    \end{align}
    Using our technique we can track the operations on the quadrature
    \begin{align}
        \hat q_1 \to \hat Q_1 &= \hat U^\dagger \hat q_1 \hat U=\hat q_1-\hat p_2.
    \end{align}
    Next, analyzing each quadrature we see that
    \begin{align}
        s_1\cos\theta_1\hat q_1-s_1\sin\theta_1\hat p_1=\hat q_1\nonumber\\
        s_2\cos\theta_2\hat q_2-s_2\sin\theta_2\hat p_2=-\hat p_2
    \end{align}
    for which we can identify
    \begin{align}
        \theta_1=0 & \quad s_1=1,\nonumber\\
        \theta_2=\pi/2 &\quad s_2=1.
    \end{align}
    From these values, we can calculate the $\Delta_j$ from Eq.~(\ref{eq:deltacasesrestricted}) as 
    \begin{align}
        \Delta_1 = 2 \quad \Delta_2 = 1.
    \end{align}
    
    This results in the following PDF:
    \begin{align}
        \label{eq:examplepdf}
        \text{PDF}(\hat Q_1=x_1)
        =&\prod_{j=1}^n
        \sum_{m_j}\delta\left(\sum_{i=1}^{n} m_i\sqrt\pi \Delta_i+c-x_1\right)\nonumber\\
        =&
        \sum_{m_1}\sum_{m_2}\delta\left(m_1\sqrt\pi \Delta_1+m_2\sqrt\pi \Delta_2-x_1\right)\nonumber\\
        =&
        \sum_{m_1}\sum_{m_2}\delta\left(2m_1\sqrt\pi+m_2\sqrt\pi-x_1\right).
    \end{align}
    
    In this example, a further simplification can be made by reparameterizing the variables ${m_2\to m_2-2m_1}$ such that
    \begin{align}
        \label{eq:examplepdf-simplified}
        \text{PDF}(\hat Q_1=x_1)
        =&
        \sum_{m_1}\delta\left(m_1\sqrt\pi-x_1\right),
    \end{align}
    which informs us that the variable $x_1$ will be measured as some randomly selected integer multiple of $\sqrt\pi$. This agrees with the qubit-based simulation at the beginning of this subsection since we measure, with equal probability, an outcome which would correspond to a $\ket{0_{\text{GKP}}}$ or  $\ket{1_{\text{GKP}}}$ state.

\subsection{Simulatable operations in single-mode measurement circuits}
\label{sec:appendix-circuit-diagrams-A}
The class $\mathcal A$ of operations, for which single-mode measurements are possible to simulate, can be interpreted in two ways. In both of these interpretations we focus on the allowed symplectic matrices because we can simulate all displacements in phase space and insert them freely without changing the structure of the symplectic matrix~\cite{serafini2017}, see Eq.~(\ref{eq:commutation-displacements}).

The first interpretation can be seen directly from the calculation of the PDF in the Heisenberg picture in Eq.~(\ref{eq:pdf-heisenbergpic-single}). Namely, we can understand the measurement mode as undergoing entangling SUM operations which transform it to $\hat q_1\to \hat q_1+\dots +\hat q_n$. This Heisenberg-evolved measurement operator is then evaluated on the rotated and squeezed GKP state. We can therefore picture the class of simulatable circuits as those whereby the input GKP states undergo rotations by angles $\theta_j\in\Theta$ and squeezing operations followed by SUM gates from each mode onto the measured mode, and by an arbitrary symplectic operation on the remaining $n-1$ modes. This is depicted in Fig.~\ref{fig:circuitclass-A-decomposition1}.

\begin{figure}[ht]
 \centering
 $$
 \Qcircuit @C=1.5em @R=1em {
 \push{\rule{1.2em}{0em}}&\lstick{\ket{0_{\text{GKP}}}} &\qw & \gate{\hat R(\theta_1)}  & \qw & \gate{\hat S(s_1)}& \qw &  \targ & \qw & \targ & \qw & \targ & \qw  &\qw&\qw & \measureD{\hat q_1} \\
 \push{\rule{1.2em}{0em}}&\lstick{\ket{0_{\text{GKP}}}} & \qw & \gate{\hat R(\theta_2)} & \qw & \gate{\hat S(s_2)}& \qw & \ctrl{-1}&\qw&\qw &\qw &\qw&\qw & \multigate{6}{\text{Sp}(2n-2,\mathbb R)} & \qw \\
 & & &\lstick{\vdots} & &\lstick{\vdots}&&&&&& &&& & \\
 & & &  & &&&&&&&&& & \\
 \rstick{\vdots}&& &  \gate{\hat R(\theta_j)}  &\qw & \gate{\hat S(s_j)}& \qw & \qw &\qw &  \ctrl{-4}&\qw&\qw&\qw& \ghost{\text{Sp}(2n-2,\mathbb R)}& \qw \\
 & & &\lstick{\vdots} & &\lstick{\vdots}&&&&&&&&&\\
 & & &  & &&&&&&&&&& \\
 \push{\rule{1.2em}{0em}}&\lstick{\ket{0_{\text{GKP}}}} & \qw &\gate{\hat R(\theta_n)}&\qw & \gate{\hat S(s_n)}& \qw & \qw &\qw &  \qw &\qw&\ctrl{-7}&\qw&\ghost{\text{Sp}(2n-2,\mathbb R)}&\qw }
 $$
 \caption{The class of simulatable operations $\mathcal A$ for single-mode measurement as understood from Eq.~(\ref{eq:pdf-heisenbergpic-single}). These are circuits initiated with $0$-logical GKP states, acted upon by single-mode rotation operations with angles $\theta_j\in\Theta$ followed by arbitrary single-mode squeezing operations and SUM operations. Following this, the first mode is measured and any additional Gaussian operation may be applied to the remainder modes. Intermediate phase space displacements may be included at any point in the circuit.}
 \label{fig:circuitclass-A-decomposition1}
\end{figure}
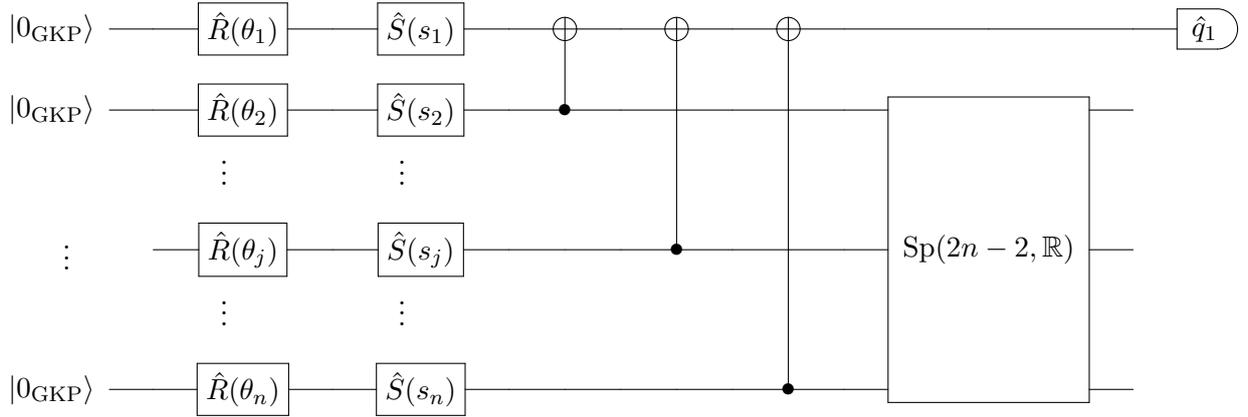

The second interpretation relies on the decomposition of the restricted set of allowed matrices given in Eq.~(\ref{eq:rspconstraints-maintext}), and on the circuit interpretation of this class that will be provided in  Appendix \ref{sec:appendix-circuit-diagrams-B}. Indeed a symplectic matrix in class $\mathcal{A}$ must have elements of $A_{1,i} = 0$  or $B_{1,i} = 0$ or $A_{1,i}/B_{1,i} =u_i/v_i$ where $u_i\in \mathbb Z$ and $v_i\in\mathbb Z_{\text{odd}}$. We can ensure that this is satisfied by using a symplectic matrix selected from the class $\mathcal B$ of simulatable operations for multimode measurement, followed by an arbitrary symplectic matrix applied to all but the measured mode. This can be seen by analyzing the symplectic matrix given in Eq.~(\ref{eq:operations-dsp}), i.e.,
\begin{align}
    \label{eq:operations-dsp-sup-calc}
    \mqty(1&0\\{\tilde C}{\tilde A}^{-T}&1)\mqty({\tilde A}&0\\0&{\tilde A}^{-T})\mqty(\text{diag}(\cos\vec \theta)&\text{diag}(\sin\vec \theta)\\
    -\text{diag}(\sin\vec \theta)&\text{diag}(\cos\vec \theta))=\mqty({\tilde A}&0\\{\tilde C}&({\tilde A}^{-T}))\mqty(\text{diag}(\cos\vec \theta)&\text{diag}(\sin\vec \theta)\\
    -\text{diag}(\sin\vec \theta)&\text{diag}(\cos\vec \theta))
\end{align}
which gives the block matrices $A=\tilde A\text{diag}(\cos\vec\theta)$ and $B=\tilde A\text{diag}(\sin\vec\theta)$. Inspecting the elements of the first row, we have
\begin{align}
    A_{1,i}=\tilde A_{1,i}\cos\theta_i \quad \text{and} \quad  B_{1,i}=\tilde A_{1,i}\sin\theta_i.
\end{align}
For a given column $i$ of the block matrices, if any of $\tilde A_{1,i}$ or $\cos\theta_i$ or $\sin\theta_i$ is zero then the $i$-th condition on the  elements of the first row of the symplectic matrix is satisfied trivially because it ensures that $A_{1,i}=0$ or $B_{1,i}=0$. Otherwise, we have
\begin{align}
    A_{1,i}/B_{1,i}=\cos\theta_i/\sin\theta_i=\cot\theta_i,
\end{align}
which will trivially satisfy the constraint if $\theta_i$ is selected from $\Theta$ in Eq.~(\ref{eq:anglesset}). A further arbitrary symplectic matrix $\text{Sp}(2(n-1),\mathbb R)$ can be applied to all but the measured mode. When applied to the symplectic matrix given in Eq.~(\ref{eq:operations-dsp-sup-calc}), it will have the effect of the identity on the measurement mode, while transforming the coefficients of the remaining modes.

The first matrix on the left hand side of Eq.~(\ref{eq:operations-dsp-sup-calc}) corresponds to the lens subgroup~\cite{arvind1995} of symplectic matrices, $\mathcal T$. The second matrix on the left-hand side of Eq.~(\ref{eq:operations-dsp-sup-calc}) corresponds to the intersection of the general linear group $\text{GL}(n,\mathbb R)$ and the subset of Hermitian positive definite symplectic matrices $\Pi(n)$ in $\text{Sp}(2n,\mathbb R)$, yielding $\text{GL}(n,\mathbb R)\cap \Pi(n)$~\cite{arvind1995}. As we show in Appendix~\ref{sec:appendix-circuit-diagrams-B}, additional squeezing operations can be added, preserving the structure of the group $\text{GL}(n,\mathbb R)\cap \Pi(n)$. We also provide a more detailed analysis of these sets of operations in Appendix~\ref{sec:appendix-circuit-diagrams-B}.
We can therefore represent the set of allowed circuits as those depicted in Fig.~\ref{fig:circuitclass-A-decomposition2}. 
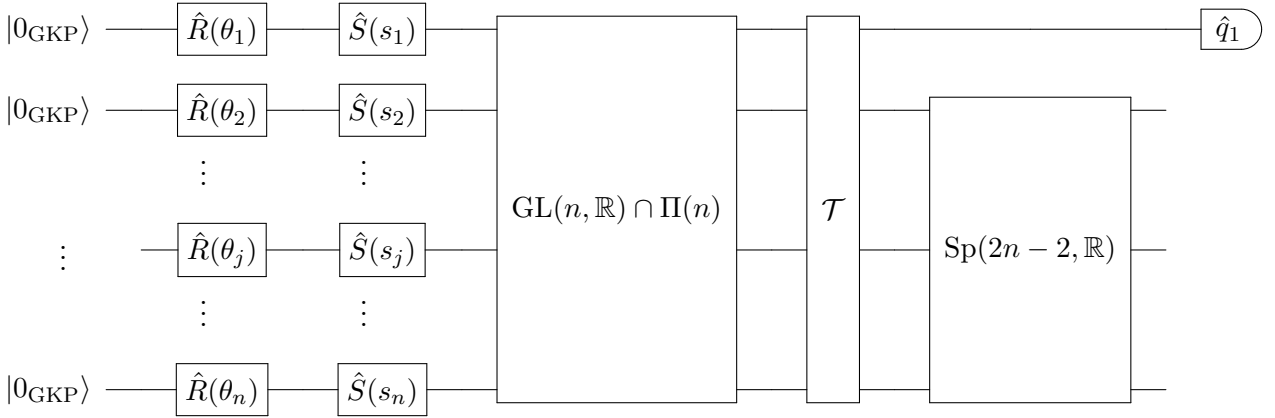
\begin{figure}[ht]
     \centering
     $$
     \Qcircuit @C=1.2em @R=1em {
     \push{\rule{1.8em}{0em}}&\lstick{\ket{0_{\text{GKP}}}} &\qw & \gate{\hat R(\theta_1)}  & \qw & \gate{\hat S(s_1)}& \qw & \multigate{7}{\text{GL}(n,\mathbb R)\cap \Pi(n)} & \qw & \multigate{7}{\mathcal T} & \qw &\qw &\qw & \measureD{\hat q_1} \\
     \push{\rule{1.8em}{0em}}&\lstick{\ket{0_{\text{GKP}}}} & \qw & \gate{\hat R(\theta_2)} & \qw & \gate{\hat S(s_2)}& \qw & \ghost{\text{GL}(n,\mathbb R)\cap \Pi(n)} & \qw & \ghost{\mathcal T} & \qw & \multigate{6}{\text{Sp}(2n-2,\mathbb R)} &\qw \\
     & & &\lstick{\vdots} & &\lstick{\vdots}&&&&&& & \\
     & & &  & &&&&&&& \\
     \rstick{\vdots}&& &  \gate{\hat R(\theta_j)}  &\qw & \gate{\hat S(s_j)}&\qw & \ghost{\text{GL}(n,\mathbb R)\cap \Pi(n)} & \qw & \ghost{\mathcal T} & \qw & \ghost{\text{Sp}(2n-2,\mathbb R)} & \qw  \\
     & & &\lstick{\vdots} & &\lstick{\vdots}&&&&&&\\
     & & &  & &&&&&&& \\
     \push{\rule{1.8em}{0em}}&\lstick{\ket{0_{\text{GKP}}}} & \qw &\gate{\hat R(\theta_n)}&\qw & \gate{\hat S(s_n)}  & \qw & \ghost{\text{GL}(n,\mathbb R)\cap \Pi(n)} & \qw & \ghost{\mathcal T} & \qw & \ghost{\text{Sp}(2n-2,\mathbb R)} & \qw }
     $$
     \caption{The class of simulatable operations $\mathcal A$ for single-mode measurement as understood from Eq.~(\ref{eq:operations-dsp-sup-calc}). The circuit has $0$-logical GKP states as input followed by single-mode rotations with rotation angles $\theta_j\in\Theta$. This is followed by arbitrary single-mode squeezing and then operations selected from $\text{GL}(n,\mathbb R)\cap \Pi(n)$ and $\mathcal T$ respectively. The single mode is measured and arbitrary symplectic operations may be performed on the remaining modes. Arbitrary displacements may be inserted at any intermediate step of the circuit.}
     \label{fig:circuitclass-A-decomposition2}
    \end{figure}
\section{Simulation of multimode measurements for Gaussian operations in $\mathcal B$}
\label{sec:appendix-multimodeoperations}

\subsection{Calculating the PDF}

\label{sec:appendix-multimodeoperations-pdf}
To calculate the PDF of measuring quadratures $\hat q_1,\dots,\hat q_n$ for the class of operations $\mathcal B$ considered in Section \ref{sec:simulation-singlemode}, we will again utilize the Heisenberg picture formalism to track the measurement operators and then find an expression for the PDF in terms of rotated single-mode GKP states.

    The PDF of the measurement in the multimode case can be evaluated analogously to in the single-mode case by inserting the identity over all rotated quadratures $\hat z^{\theta_i}_i$,
    \begin{align}
        \mathbbm 1 = \prod_{j=1}^n \int \dd z_j \ket{\hat z_j^{\theta_j}=z_j}\bra{\hat z_j^{\theta_j}=z_j}= \int \dd \mathbf z \prod_{j=1}^n  \ket{\hat z_j^{\theta_j}=z_j}\bra{\hat z_j^{\theta_j}=z_j}.
    \end{align}
    The PDF can then be evaluated in terms of the Heisenberg evolved measurement operators Eq.~(\ref{eq:heisenberg-opterator-multi}) as
    \begin{align}
     &\text{PDF}(\mathbf x)\nonumber \\
     =&\bra{0_{\text{GKP}}}\dots \bra{0_{\text{GKP}}} \prod_{j=1}\ket{\hat Q_j=x_j}\bra{\hat Q_j=x_j}\ket{0_{\text{GKP}}}\dots \ket{0_{\text{GKP}}}\nonumber \\
     =&\bra{0_{\text{GKP}}}\dots \bra{0_{\text{GKP}}} \prod_{j=1}\delta(\hat Q_j-x_j)\ket{0_{\text{GKP}}}\dots \ket{0_{\text{GKP}}}\nonumber \\
    =&\int \dd \mathbf z \bra{0_{\text{GKP}}}\dots \bra{0_{\text{GKP}}} \prod_{j=1}\delta(\hat Q_j-x_j)\prod_i\left(\ket{\hat z_j^{\theta_j}=z_j}\bra{\hat z_j^{\theta_j}=z_j}\right)\ket{0_{\text{GKP}}}\dots \ket{0_{\text{GKP}}}\nonumber \\
    =&\int \dd \mathbf z \bra{0_{\text{GKP}}}\dots \bra{0_{\text{GKP}}} \prod_{j=1}\delta\left(\sum_ia^{(j)}_i\hat r^{\theta_i}_{i}+c_j-x_j\right)\prod_i\left(\ket{\hat z_j^{\theta_j}=z_j}\bra{\hat z_j^{\theta_j}=z_j}\right)\ket{0_{\text{GKP}}}\dots \ket{0_{\text{GKP}}}\nonumber \\
    =&\int \dd \mathbf z \bra{0_{\text{GKP}}}\dots \bra{0_{\text{GKP}}} \prod_{j=1}\delta\left(\sum_ia^{(j)}_iz_i+c_j-x_j\right)\prod_i\left(\ket{\hat z_j^{\theta_j}=z_j}\bra{\hat z_j^{\theta_j}=z_j}\right)\ket{0_{\text{GKP}}}\dots \ket{0_{\text{GKP}}}\nonumber \\
    =&\int \dd \mathbf z \prod_{j=1}\left(\delta\left(\sum_ia^{(j)}_iz_i+c_j-x_j\right)\right)\prod_i\abs{\bra{0_{\text{GKP}}}\ket{\hat z_i^{\theta_i}=z_i}}^2.
\end{align}
We can then use
\begin{align}
\label{eq:finalprobsinglemode-z}
    \abs{\bra{0_{\text{GKP}}}\ket{\hat z_i^{\theta_i}=z_i}}^2=\abs{\psi_{\theta}(z_i)}^2\propto&\sum_{m\in\mathbb Z} \delta(z_i-m\sqrt\pi \Delta_i)
\end{align}
to get
\begin{align}
     \text{PDF}(\mathbf x)=&\int \dd \mathbf z \prod_{j=1}^n\left(\delta\left(\sum_{i=1}^na^{(j)}_iz_i+c_j-x_j\right)\right)\prod_{k=1}^n\sum_{m_k\in\mathbb Z} \delta(z_k-m_k\sqrt\pi \Delta_k)\nonumber\\
     =&\int \dd \mathbf z \left(\delta\left(\sum_{i=1}^na^{(1)}_iz_i+c_1-x_1\right)\dots \delta\left(\sum_{i=1}^na^{(n)}_iz_i+c_n-x_n\right)\right)\prod_{k=1}^n\sum_{m_k\in\mathbb Z} \delta(z_k-m_k\sqrt\pi \Delta_k)\nonumber\\
     =&\sum_{m_1,\dots,m_n\in\mathbb Z}\left(\delta(\sum_{i=1}^na_i^{(1)}m_i\sqrt\pi\Delta_i+c_1-x_1)\right)\dots \left(\delta(\sum_{i=1}^na_i^{(n)}m_i\sqrt\pi\Delta_i+c_n-x_n)\right)\nonumber\\
     =&\sum_{m_1,\dots,m_n\in\mathbb Z}
     \prod_{j=1}^n\delta(\sum_{i=1}^na_i^{(j)}m_i\sqrt\pi\Delta_i+c_j-x_j).
     \label{eq:appendix-multimode-pdf}
\end{align}

    We again have the $\Delta_j$ specified by Eq.~(\ref{eq:deltacases}) for each mode, i.e.
    \begin{align}
    \label{eq:deltacasesrestricted-dsp}
        \Delta_j=\begin{cases}
        \sin(\theta_j) /v_j\quad & \text{ (case 1) if } \cot\theta_j=u_j/v_j : u\in\mathbb Z,v_j\in\mathbb Z_{\text{odd}} \text{ and } \gcd(u_j,v_j)=1,\\
        2\quad &\text{ (case 2) if } \theta_j=k\pi \text{ for } k \in \mathbb Z.
        \end{cases}
    \end{align}

\subsection{Simple Example}
\label{sec:appendix-multimodeoperations-example}
    In this section, we expand the simple example of the previous appendix to illustrate how our simulation technique works for multimode measurement.
    We would again like to simulate the following operations
    \begin{align}
        \label{eq:simple-example-multi-op}
        \hat U=e^{-i\hat q_1\hat p_2}\hat F_1
    \end{align}
    acting on GKP input states. Note that the operation in Eq.~(\ref{eq:simple-example-multi-op}) belongs to classes $\mathcal A$ and $\mathcal B$ simultaneously.
    
    Using our technique we can again track the Heisenberg evolution of the measurement operators, however, this time we will track both quadratures:
    \begin{align}
        \hat q_1 \to & \hat Q_1 = \hat U^\dagger \hat q_1 \hat U=-\hat p_1\\
        \hat q_2 \to & \hat Q_2 = \hat U^\dagger \hat q_2 \hat U=-\hat p_1 +\hat q_2.
    \end{align}
    
    The symplectic matrix of this operation can be represented as a rotation of the first quadrature, followed by a mode mixing operation. The rotation is parametarized by $\vec\theta=(\pi/2,0)$ for which we can build the matrix
    \begin{align}
        \mqty(\text{diag}(\cos\vec \theta)&\text{diag}(\sin\vec \theta)\\
    -\text{diag}(\sin\vec \theta)&\text{diag}(\cos\vec \theta))
    \end{align}
    with $\text{diag}(\cos\vec\theta)=\text{diag}(0,1)$ and $\text{diag}(\sin\vec\theta)=\text{diag}(1,0)$.
    The mode mixing operation is the $C_X$ operation which is in the block diagonal form
    \begin{align}
        \mqty(A&0\\0&(A^T)^{-1})
    \end{align}
    with
    \begin{align}
        A=\mqty(1&0\\1&1).
    \end{align}
    Therefore, we can evaluate the PDF, which is given by Eq.~(\ref{eq:appendix-multimode-pdf}) as
    \begin{align}
         \text{PDF}(\mathbf x)=&\sum_{m_1,\dots,m_n\in\mathbb Z}\prod_{j=1}^n\delta(\sum_{i=1}^na_i^{(j)}m_i\sqrt\pi\Delta_i+c_j-x_j)
    \end{align}
    where we evaluate the parameters $\Delta_i$ according to Eq.~(\ref{eq:deltacasesrestricted-dsp}) as
    \begin{align}
        \Delta_1=&\sin(\pi/2)/1=1\nonumber\\
        \Delta_2=&2.
    \end{align}
    Therefore, the PDF can be written explicitly as
    \begin{align}
         \text{PDF}(\mathbf x)=&\sum_{m_1,\dots,m_n\in\mathbb Z}
     \prod_{j=1}^n\delta(\sum_{i=1}^na_i^{(j)}m_i\sqrt\pi\Delta_i+c_j-x_j)\nonumber\\
     =&\sum_{m_1,m_2\in\mathbb Z}
     \delta(\sum_{i=1}^na_i^{(1)}m_i\sqrt\pi\Delta_i+c_1-x_1)\delta(\sum_{i=1}^na_i^{(2)}m_i\sqrt\pi\Delta_i+c_2-x_2) \nonumber\\
     =&\sum_{m_1,m_2\in\mathbb Z}
     \delta(a_1^{(1)}m_1\sqrt\pi\Delta_1+a_2^{(1)}m_2\sqrt\pi\Delta_2+c_1-x_1)\delta(a_1^{(2)}m_1\sqrt\pi\Delta_1+a_2^{(2)}m_2\sqrt\pi\Delta_2+c_2-x_2)\nonumber \\
     =&\sum_{m_1,m_2\in\mathbb Z}
     \delta(m_1\sqrt\pi-x_1)\delta(m_1\sqrt\pi+2m_2\sqrt\pi-x_2).
    \end{align}
\subsection{Simulatable operations in multimode measurement circuits}
\label{sec:appendix-circuit-diagrams-B}
In this subsection, we express the allowed operations in circuit class $\mathcal B$ in terms of circuit diagrams. Similarly to Appendix \ref{sec:appendix-circuit-diagrams-A} we can analyze the decomposition of the allowed symplectic matrix. Note that we can freely insert displacements in phase space anywhere in the circuit, as these preserve the structure of the symplectic matrix, as in Eq.~(\ref{eq:commutation-displacements})~\cite{serafini2017}.

The class $\mathcal B$ is composed of symplectic operations which can be decomposed as 
\begin{align}
    \label{eq:operations-dsp-sup}
    \mqty(1&0\\{\tilde C}{\tilde A}^{-T}&1)\mqty({\tilde A}&0\\0&{\tilde A}^{-T})\mqty(\text{diag}(\cos\vec \theta)&\text{diag}(\sin\vec \theta)\\
    -\text{diag}(\sin\vec \theta)&\text{diag}(\cos\vec \theta)).
\end{align}
where the angles $\theta_i\in \Theta$, as defined in Eq.~(\ref{eq:anglesset}).

The first matrix
\begin{align}
    \mqty(1&0\\{\tilde C}{\tilde A}^{-T}&1)
\end{align}
corresponds to the lens subgroup~\cite{arvind1995} of symplectic matrices, $\mathcal T$. These are transformations which take
\begin{align}
    \hat{\mathbf q} &\to \hat{\mathbf q}\\
    \hat{\mathbf p} &\to \hat{\mathbf p}+{\tilde C}({\tilde A}^{T})^{-1}\hat{\mathbf q}.
\end{align}
    
The matrices of the form
    \begin{align}
        \mqty({\tilde A}&0\\0&({\tilde A}^{T})^{-1})
    \end{align}
    correspond to the intersection of the general linear group $\text{GL}(n,\mathbb R)$ and the subset of Hermitian positive definite symplectic matrices $\Pi(n)$ in $\text{Sp}(2n,\mathbb R)$, yielding $\text{GL}(n,\mathbb R)\cap \Pi(n)$~\cite{arvind1995}. This describes transformations of the form
    \begin{align}
        \hat{\mathbf q} &\to {\tilde A}\hat{\mathbf q}\\
        \hat{\mathbf p} &\to {\tilde A}^{-1}\hat{\mathbf p}
    \end{align}
    where $\tilde A$ is symmetric.
    We are unaware of a method to decompose this set of operations into single and two-mode operations. However, we can immediately notice that combining two operations of this form in sequence will result in a new matrix which also conforms to this structure. Formally this can be expressed as
    \begin{align}
    \label{eq:squeezing-free}
    \mqty({\tilde A}&0\\0&({\tilde A}^{T})^{-1})\mqty({\tilde A'}&0\\0&({\tilde A}^{\prime T})^{-1}) \in \text{GL}(n,\mathbb R)\cap \Pi(n) \nonumber \\
    \forall \quad  \mqty({\tilde A}&0\\0&({\tilde A}^{T})^{-1}) \text{ and } \mqty({\tilde A'}&0\\0&({\tilde A}^{\prime T})^{-1}) \in \text{GL}(n,\mathbb R)\cap \Pi(n)
    \end{align}
    which can be confirmed by analyzing the product of two arbitrary matrices where $\tilde A,\tilde A'$ are non-singular $n\times n$ matrices:
    \begin{align}
        \mqty({\tilde A}&0\\0&({\tilde A}^{T})^{-1})\mqty({\tilde A'}&0\\0&({\tilde A}^{\prime T})^{-1})=\mqty({\tilde A}{\tilde A'}&0\\0&({\tilde A}^{T})^{-1}({\tilde A}^{\prime T})^{-1})=\mqty({\tilde A}{\tilde A'}&0\\0&(\tilde A\tilde A')^{-T}).
    \end{align}
    The set of single-mode squeezing operations can be expressed as
    \begin{align}
        \mqty(\text{diag}(\vec s)&0\\0&\text{diag}(\vec s^+))=\mqty(\text{diag}(\vec s)&0\\0&\text{diag}(\vec s)^{-1})
    \end{align}
    where $\vec s$ is a vector of non-zero values of squeezing and $\vec s^+$ is the element-wise inverse of $\vec s$. We can therefore combine single-mode squeezing operations with any operation selected from $\text{GL}(n,\mathbb R)\cap \Pi(n) $ while maintaining that the corresponding symplectic matrix belongs to the same set.
    
  We can therefore consider that the simulatable circuits in class $\mathcal B$ are those for which there exists a circuit of the form shown in Fig.~\ref{fig:circuitclass-B-decomposition}.   Comparing this figure with Fig.~\ref{fig:circuitclass-A-decomposition2} makes clear the interpretation of the class of circuits corresponding to the larger set $\mathcal{A}$ as the same operations in $\mathcal{B}$, followed by arbitrary symplectic operations on the last $n - 1$ modes.
    
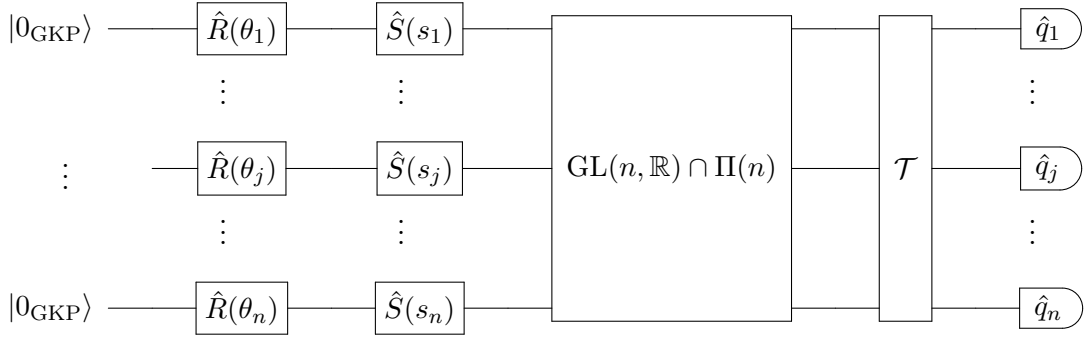
\begin{figure}[ht]
     \centering
     $$
     \Qcircuit @C=1.5em @R=1em {
     \push{\rule{1.2em}{0em}}&\lstick{\ket{0_{\text{GKP}}}} &\qw & \gate{\hat R(\theta_1)} & \qw & \gate{\hat S(s_1)} & \qw & \multigate{6}{\text{GL}(n,\mathbb R)\cap \Pi(n)} & \qw & \multigate{6}{\mathcal T} & \qw & \measureD{\hat q_1}\\
     & & &\lstick{\vdots} & &\lstick{\vdots}&&&&&&\lstick{\vdots}\\
     & & &  & &&&&&&& \\
     \rstick{\vdots}&& & \gate{\hat R(\theta_j)} &\qw &\gate{\hat S(s_j)}  &\qw & \ghost{\text{GL}(n,\mathbb R)\cap \Pi(n)} & \qw & \ghost{\mathcal T} & \qw & \measureD{\hat q_j}  \\
     & & &\lstick{\vdots} & &\lstick{\vdots}&&&&&&\lstick{\vdots}\\
     & & &  & &&&&&&& \\
     \push{\rule{1.2em}{0em}}&\lstick{\ket{0_{\text{GKP}}}} & \qw &\gate{\hat R(\theta_n)}&\qw & \gate{\hat S(s_n)}  & \qw & \ghost{\text{GL}(n,\mathbb R)\cap \Pi(n)} & \qw & \ghost{\mathcal T} & \qw & \measureD{\hat q_n}}
     $$
     \caption{The circuit diagram of arbitrary circuits selected from the simulatable class $\mathcal B$. The first action of the circuit on the modes is a rotation of each mode with angles $\theta_i\in\Theta$ as defined in Eq.~(\ref{eq:anglesset}), which corresponds to the third matrix in Eq.~(\ref{eq:operations-dsp-sup}). This is followed by arbitrary squeezing of each mode, and can be trivially included as a result of Eq.~(\ref{eq:squeezing-free}). The next operation is selected from $\text{GL}(n,\mathbb R)\cap \Pi(n) $ which corresponds to the second matrix in Eq.~(\ref{eq:operations-dsp-sup}). The final operation is selected from $\mathcal T$ and corresponds to the third matrix in Eq.~(\ref{eq:operations-dsp-sup}). Finally, the modes are measured in the position basis with homodyne detection. We note that arbitrary phase space displacements can be inserted into the circuit due to the fact that they preserve the structure of the symplectic matrix, in accordance with Eq.~(\ref{eq:commutation-displacements}).}
     \label{fig:circuitclass-B-decomposition}
    \end{figure}

\section{Density of the set of simulatable operations in class $\mathcal A$}
\label{sec:appendix-restrictedsets-density}
        In this Appendix, we will demonstrate that the circuit class $\mathcal A$ contains symplectic matrices which have parameters which are dense in the reals.
        Using the decomposition in Section \ref{sec:tracking-evolution} we have shown that we can construct a symplectic matrix $M$ with a free choice parameters ${A_{1,i}=a_i\in\mathbb R}$ and ${B_{1,i}=b_i\in\mathbb R}$, for $i\in\{1,\dots,n\}$. However, in order to simulate the measurement of the output mode we must make a further restriction on these variables. We know that for a symplectic matrix to exist in $\text{RSp}(2n,\mathbb R)$ it must have either one of $A_{1,i}=0$ or $B_{1,i}=0$ or $A_{1,i}/B_{1,i}\in \mathbb Q_{(2)}$.
        
        By introducing a new set
    \begin{align}
    \label{eq:setoddoddzero}
        \mathbb Q_{\text{odd}}=\{u/v : u\in\mathbb Z_{\text{odd}},v\in\mathbb Z_{\text{odd}}\}
    \end{align}
    which is contained by $\mathbb Q_{(2)}$ but also contains fewer elements that $\mathbb Q_{(2)}$, we can create arbitrary symplectic matrices which will always be contained within $\text{RSp}(2n,\mathbb R)$.
    If we select the matrix elements $a_i,b_i\in\mathbb Q_{\text{odd}}$ we can guarantee that the matrix will be contained in $\text{RSp}(2n,\mathbb R)$, since the division of one element in $\mathbb Q_{\text{odd}}$ by another element in $\mathbb Q_{\text{odd}}$, will always give an element in  $\mathbb Q_{(2)}$. Next, we can show that the set from which we choose the matrix elements $\mathbb Q_{\text{odd}}$ is dense on the reals. 

    Formally we can prove this following the technique of Ref.~\cite{trench2003} to prove the density of the rational numbers on the reals. The density of a set $\mathcal \chi$ on the reals formally means that for any $x,y\in\mathbb R$ where $x<y$ then there exists $\alpha\in \mathcal \chi$ such that $x<\alpha<y$. 
    
    First we recall the demonstration that the rationals are dense on the reals, following the proof given by Ref.~\cite{trench2003}. This means that for any $x<y$ with $x,y\in\mathbb R$ we must have some $q\in\mathbb Q$ such that $x<q<y$.

    By the Archimedian property~\cite{trench2003}, for any $x,y$ with $x<y$ there exists some integer $\beta\in\mathbb N$ such that $0<1/\beta<(y-x)$.  There must also exist some integer $\alpha$ such that $\alpha>\beta x$. Let $\alpha$ be the smallest integer such that $\alpha>\beta x$, which implies $\alpha-1\le \beta x$. This leads us to the relation
    \begin{align}
        \beta x<\alpha<\beta+1.
    \end{align}
    We know that $1<\beta (y-x)$ and therefore
    \begin{align}
        \beta x<\alpha<\beta x+\beta (y-x)=\beta y
        \implies & x<\alpha/\beta<y.
    \end{align}
    This means that for any $x<y$ such that $x,y\in\mathbb R$ we can define $\alpha\in\mathbb Z$ and $\beta\in\mathbb N$, where $\mathbb N=\{1,2,3,\dots \}$ are the natural numbers, not including $0$.
    
    Now we will follow the same proof technique to show that the set $\mathbb Q_{\text{odd}}$ is dense on the reals. Again, by the Archimedian property, there exists some $\beta \in\mathbb N$ s.t. $0<3/(2\beta+1)<(y-x)$. There must also exist some integer $\alpha$ such that ${(2\alpha+1)>(2\beta+1)x}$. Let $\alpha$ be the smallest integer such that ${(2\alpha+1)>(2\beta+1)x}$, which implies ${2\alpha-1\le(2\beta+1)x\implies 2\alpha\le(2\beta+1)x+1\implies2\alpha<(2\beta+1)x+2}$. This leads us to the relation
    \begin{align}
        (2\beta+1)x<2\alpha+1<(2\beta+1)x+3.
    \end{align}
    We know that $3<(2\beta+1)(y-x)$ and therefore
    \begin{align}
        (2\beta+1)x<2\alpha+1<(2\beta+1)x+(2\beta+1)(y-x)=(2\beta+1)y.
    \end{align}
    This means that for any $x<y$ such that $x,y\in\mathbb R$ we can define $\alpha\in\mathbb Z$ and $\beta\in\mathbb N$ such that
    \begin{align}
        x<\frac{2\alpha+1}{2\beta+1}<y
    \end{align}
    i.e. there exists a number $r\in \mathbb Q_{\text{odd}}$ such that $x<r<y$.
    
    Therefore, for any two symplectic matrices which have the effect on the output mode as
    \begin{align}
        M: \hat q_1 \to a_1 \hat q_1 + b_1\hat p_1 +\dots +a_n \hat q_n +b_n \hat p_n
    \end{align}
    and
    \begin{align}
        M': \hat q_1 \to a_1' \hat q_1 + b_1'\hat p_1 +\dots +a_n' \hat q_n +b_n' \hat p_n
    \end{align}
    where $a_i,b_i,a'_i,b'_i\in\mathbb R$, it is always possible to select parameters $\bar a_i,\bar b_i\in\mathbb Q_{\text{odd}}$ such that
    \begin{align}
        a_i<\bar a_i<a_i' \quad \text{ or } \quad a_i'<\bar a_i<a_i,\quad \text{and} \quad b_i<\bar b_i<b_i' \quad \text{ or } \quad b_i'<\bar b_i<b_i  \quad \forall i\in\mathbb Z_n,
    \end{align}
    whereby
    \begin{align}
        \bar M: \hat q_1 \to \bar a_1 \hat q_1 + \bar b_1\hat p_1 +\dots +\bar a_n \hat q_n +\bar b_n \hat p_n
    \end{align}
    is simulatable by our technique.

\section{The Clifford group is not contained in the set of simulatable operations}
\label{sec:appendix-restrcitedsets-clifford}
We show using a simple example that the Clifford group is not contained within $\text{RSp}(2n,\mathbb R)$. We will use the Fourier transform $\hat F$ which transforms the quadratures according to Eq.~(\ref{eq:fourier-effects}). We will also use the phase gate $\hat P_1=e^{i\hat q_1^2/2}$~\cite{gottesman2001} which transforms the quadratures as
\begin{align}
    \hat q_1 &\to \hat q_1 \nonumber \\
    \hat p_1 &\to \hat q_1+\hat p_1.
\end{align}
We can define a single qubit Clifford operation in the GKP encoding as $\hat U=\hat F_1\hat P_1^2\hat F_1$ and consider its effect on the Heisenberg measurement operators as
\begin{align}
    \hat Q_1=\hat U\hat q_1\hat U^\dagger=&\hat F_1\hat P_1\hat P_1\hat F_1\hat q_1\hat F_1^\dagger \hat P_1^\dagger\hat P_1^\dagger\hat F_1^\dagger\\
    =&\hat F_1\hat P_1\hat P_1(-\hat p_1)\hat P_1^\dagger\hat P_1^\dagger\hat F_1^\dagger \\
    =&\hat F_1(-(2\hat q_1+\hat p_1))\hat F_1^\dagger  \\
    =&-(-2\hat p_1+\hat q_1)\\
    =&2\hat p_1-\hat q_1.
\end{align}

The top blocks of the $2\times 2$ symplectic matrix $M$ corresponding to the operation $\hat U$ can be written as
\begin{align}
    \mqty(-1&2).
\end{align}
We know that a symplectic matrix $M$ exists in $\text{RSp}(2n,\mathbb R)$ if and only if
\begin{align}
    M_{1,i}=0 \text{ or } M_{1,n+i}=0 \text{ or } M_{1,i}/M_{1,n+i}\in\mathbb Q_{(2)} \forall i\in \{1,\dots, n\}
\end{align}
which we can check for the symplectic matrix $M$. For a $2\times 2$ symplectic matrix, this amounts to a single condition, i.e. that
\begin{align}
    M_{1,1}=0 \text{ or } M_{1,2}=0 \text{ or } M_{1,1}/M_{1,2}\in\mathbb Q_{(2)}.
\end{align}

However, we know that $\mathbb Q_{(2)}$ is defined as the rationals such that the denominator is odd in its simplest form. The fraction $-1/2$ is already in its simplest form and has even denominator. Therefore $-1/2\notin \mathbb Q_{(2)}$, which means it is not possible to simulate the circuit using our method. We have provided an example of a Clifford circuit which does not exist within the set $\text{RSp}(2n,\mathbb R)$ and hence it also cannot be contained within $\text{DSp}(2n,\mathbb R)$.

Note, however, that the simulatability of Clifford operations was proven in Ref.~\cite{garcia-alvarez2020}.

    \section{Route to extending simulatability to realistic GKP states}
    \label{appendix:route-to-realistic}
    The circuit classes we have shown to be simulatable use ideal GKP states. Formally, these are not normalizable and therefore are also not physical states. Tackling the extension of our results on the simulatability of Gaussian circuits to the case of realistic GKP states is challenging. In this Appendix, we will sketch the calculations required to extend our result to include realistic GKP states. We first demonstrate that it is possible to express a rotated and squeezed GKP state in an analytic form using the Jacobi theta function, given in Eq.~(\ref{eq:jacobi}). We then point to the difficulty of extending our results to the case of realistic GKP states, by providing the unsolved equation of the PDF of a circuit involving multiple realistic GKP states, and showing that it cannot be easily solved in a closed form.

    \subsection{PDF of a rotated and squeezed realistic GKP state}
    
    We can evaluate the PDF of a rotated and squeezed realistic GKP state using similar techniques to the case of the ideal GKP states. We know that it is possible to write the wave function of the realistic GKP states in the form~\cite{matsuura2020,pantaleoni2021}
    \begin{align}
        \psi_{\text{GKP}}^\Delta (x)=e^{-x^2\Delta^2/2}\vartheta(\zeta=x/2\sqrt\pi,\tau=i\Delta^2/2)
    \end{align}
    where $\vartheta$ is the Jacobi theta function given in Eq.~(\ref{eq:jacobi}) and $\Delta$ is the squeezing parameter of the peaks and envelope of the GKP state.
    
    Applying rotation and squeezing to this wave function gives
    \begin{align}
        \psi^\Delta_{\theta,s}(x)&=\bra{q=x}\hat S(s)
        \hat R(\theta)\ket{0^\Delta_{\text{GKP}}}\nonumber\\
        &=\int \dd x' \bra{q=x}\hat S(s)
        \hat R(\theta)\ket{ q=x'}\bra{ q=x'}\ket{0^\Delta_{\text{GKP}}}\nonumber\\
        &=\int \dd x' \bra{q=x/s}
        \hat R(\theta)\ket{ q=x'}\bra{ q=x'}\ket{0^\Delta_{\text{GKP}}}.
    \end{align}
    We can use the propagator, which is also used in the case of infinite squeezing, Eq.~(\ref{eq:propogator}), 
    \begin{align}
        \label{eq:propogator-again}
        K(x,x';\theta)=\bra{q=x}\hat R(\theta)\ket{q=x'}=\frac{1}{\sqrt{2\pi i \sin\theta}}\exp(\frac{i}{2}\left((x-x'\sec\theta)^2\cot\theta-\tan\theta x'^2\right))
    \end{align}
    to write
    \begin{align}
        \psi^\Delta_{\theta,s}(x)&=\int \dd x' \bra{q=x}\hat S(s)
        \hat R(\theta)\ket{ q=x'}\bra{ q=x'}\ket{0^\Delta_{\text{GKP}}}\nonumber\\
        &=\int \dd x' \frac{1}{\sqrt{2\pi i \sin\theta}}e^{\frac{i}{2}\left((x/s-x'\sec\theta)^2\cot\theta-\tan\theta x'^2\right)}\bra{ q=x'}\ket{0^\Delta_{\text{GKP}}}\nonumber\\
        &=\int \dd x' \frac{1}{\sqrt{2\pi i \sin\theta}}e^{\frac{i}{2}\left((x/s-x'\sec\theta)^2\cot\theta-\tan\theta x'^2\right)}e^{-x'^2\Delta^2/2}\vartheta(\zeta=\frac{x'}{2\sqrt\pi},\tau=\frac{i\Delta^2}{2})\nonumber\\
        &=\int \dd x' \frac{1}{\sqrt{2\pi i \sin\theta}}e^{\frac{i}{2}\left((x/s-x'\sec\theta)^2\cot\theta-\tan\theta x'^2\right)}e^{-x'^2\Delta^2/2}
        \sum_n e^{-\pi  n^2 \Delta^2/2}e^{2 \pi i n x'/2\sqrt\pi}\nonumber\\
        &\propto e^{\frac{i x^2(i\Delta^2 +(1+\Delta^4)\cot\theta)}{2 s^2(1+\Delta^4-i\Delta^2\cot\theta)}}\vartheta(\zeta=-\frac{x\csc\theta}{\sqrt\pi s(1+\Delta^4 -i\Delta^2 \cot\theta)},\tau=\frac{2i(\Delta^2-i\cot\theta)}{1+\Delta^4-i\Delta^2\cot\theta}) \nonumber\\
        &= e^{x^2\gamma_{\Delta,s,\theta}}\vartheta(\zeta=x\eta_{\Delta,s,\theta}, \tau =\tau_{\Delta,\theta})
    \end{align}
    which is defined in terms of the constants
    \begin{align}
        \eta_{\Delta,s,\theta}=&-\csc\theta/\left(\sqrt\pi (s+\Delta^4 s-i\Delta^2 s\cot\theta)\right),\\
        \tau_{\Delta,\theta}=&2i(\Delta^2-i\cot\theta)/(1+\Delta^4-i\Delta^2\cot\theta), \\
        \gamma_{\Delta,s,\theta}=&\frac{i(i\Delta^2 +(1+\Delta^4)\cot\theta)}{2 s^2(1+\Delta^4-i\Delta^2\cot\theta)}.
    \end{align}
    For convenience, we drop the indices in the notation of these constants. From the wave function, we can determine that the PDF is
    \begin{align}
    \label{eq:pdf-realistic-rotated}
        |\psi^\Delta_{\theta,s}(x)|^2
        &\propto e^{x^2(\gamma+\gamma^*)}|\vartheta(\zeta=x\eta,\tau)|^2,
    \end{align}
    where we have left out the normalization constant for brevity.
    Expanding these variables in series and assuming $\Delta^4 \to 0$, we have
    \begin{align}
        \eta^{(3)}=&-\frac{\csc\theta}{\sqrt\pi  s}(1+i\cot\theta \Delta^2),\\
        \tau^{(3)}=&2\cot\theta +2i\csc^2\theta \Delta^2,\\
        \gamma^{(3)}=&\frac{i\cot\theta}{2s^2}-\frac{\csc^2\theta \Delta^2 }{2s^2},
    \end{align}
    which somewhat simplifies the PDF to
    \begin{align}
        |\psi^\Delta_{\theta,s}(x)|^2
        &\approx e^{-x^2\csc^2\theta\Delta^2/s^2}|\vartheta(\zeta=-x\frac{\csc\theta}{\sqrt\pi  s}(1+i\cot\theta \Delta^2),\tau=2\cot\theta +2i\csc^2\theta \Delta^2)|^2.
    \end{align}
    However, even with this approximation, it remains challenging to evaluate the PDF of the circuit even in the case of single-mode measurement.
    
    Note that in the limit $\Delta=0$ we can identify that these parameters will reach
    \begin{align}
        \eta^{(3)}_{\Delta \to 0}=&-\frac{\csc\theta}{\sqrt\pi  s},\\
        \tau^{(3)}_{\Delta \to 0}=&2\cot\theta,\\
        \gamma^{(3)}_{\Delta \to 0}=&\frac{i\cot\theta}{2s^2},
    \end{align}
    which returns the infinite-squeezing GKP state PDF given in Eq.~(\ref{eq:probtheta-original}),
    \begin{align}
        |\psi^{\Delta\to 0}_{\theta,s}(x)|^2&\propto|\vartheta(\zeta=-x\frac{\csc\theta}{\sqrt\pi s},\tau=2\cot\theta)|^2.
    \end{align}
    
    \subsection{Single-mode measurement}
    We showcase the complexity of evaluating the PDF of a single-mode measurement result by providing an example of a calculation involving two modes. The PDF is given by
    \begin{align}
        \text{PDF}(\hat Q_1=x_1)
        =& \int \dd z_1\dd z_2 \delta\left(s_1 z_1+s_2 z_2+c-x_1\right)\abs{\bra{0_{\text{GKP}}} \ket{\hat z_1^{\theta_1}=z_1}}^2\abs{\bra{0_{\text{GKP}}} \ket{\hat z_2^{\theta_2}=z_2}}^2 \nonumber\\
        =& \int \dd z_1\dd z_2 \delta\left((s_1 z_1+c-x_1)/s_2+ z_2\right)\abs{\bra{0_{\text{GKP}}} \ket{\hat z_1^{\theta_1}=z_1}}^2\abs{\bra{0_{\text{GKP}}} \ket{\hat z_2^{\theta_2}=z_2}}^2 \nonumber\\
        =& \int \dd z_1 \abs{\bra{0_{\text{GKP}}} \ket{\hat z_1^{\theta_1}=z_1}}^2\abs{\bra{0_{\text{GKP}}} \ket{\hat z_2^{\theta_2}=(x_1-s_1 z_1-c)/s_2}}^2.
    \end{align}
    Note that in the case of infinite squeezing, the PDF of the rotated and squeezed GKP state is a Dirac comb, i.e. Eq.~(\ref{eq:singlemodepdf-main}), which is a summation over delta functions. When inserted into the multimode PDF, the latter yields a product of summations of delta functions in terms of $z_1,\dots,z_n$, which results in a function that can be integrated over these variables, leading to Eq.~(\ref{eq:multimode-singlemeasure-pdf-main}). However, in the case of finite-squeezing, the rotated and squeezed single-mode PDF cannot be expressed in terms of delta functions, which limits the possibility of analytically solving this expression.
    Using the fact that the PDF of the rotated and squeezed GKP state is given by Eq.~(\ref{eq:pdf-realistic-rotated}), we have
    \begin{align}
        \text{PDF}(\hat Q_1=x_1)
        =& \int \dd z_1 e^{z_1^2(\gamma+\gamma^*)}|\vartheta(\zeta=z_1\eta,\tau)|^2
        e^{((x_1-s_1 z_1-c)/s_2)^2(\gamma+\gamma^*)}|\vartheta(\zeta=(x_1-s_1 z_1-c)/s_2\eta,\tau)|^2.
    \end{align}
    We are not aware of any theorem which rules out the possibility to evaluate such a function in general. However, we are also unable to provide a general analytic solution of this integral.
\end{widetext}
\clearpage \mbox{} \clearpage
\end{document}